\documentclass[
 aps,
 ,twocolumn
]{revtex4}
\usepackage{graphicx}
\usepackage{amsmath}
\usepackage{mathrsfs}

\usepackage{tikz}
\usepackage{pgflibraryarrows}
\usepackage{pgflibrarysnakes}
\usetikzlibrary{decorations.pathmorphing}
\usepgflibrary{arrows.meta}

\DeclareMathOperator{\csch}{csch}

\usepackage{subfigure}

\usepackage{braket}

\begin{document}

\title{Particle production and apparent decoherence due to an accelerated time-delay}

\author{Sho Onoe}
\email{sho.onoe@uqconnect.edu.au}
\author{Daiqin Su}
\email{sudaiqin@gmail.com}
\author{Timothy.~C.~Ralph}\email{ralph@physics.uq.edu.au}
\affiliation{Centre for Quantum Computation and Communication Technology, School of Mathematics and Physics, The University of Queensland, St. Lucia, Queensland, 4072, Australia}

\date{\today}

\begin{abstract}
{We study the radiation produced by an accelerated time-delay acting on the left moving modes. Through analysis via the Schr\"odinger picture, we find that the final state is a two-mode squeezed state of the left moving Unruh modes, implying particle production. We analyse the system from an operational point of view via the use of self-homodyne detection with broad-band inertial detectors. We obtain semi-analytical solutions that show that the radiation appears decohered when such an inertial observer analyses the information of the radiation from the accelerated time-delay source. We make connection with the case of the accelerated mirror. We investigate the operational conditions under which the signal observed by the inertial observer can be purified.
}
\end{abstract}

\maketitle


\section{Introduction}

Since the 1970s it has been well known that a moving mirror can radiate particles \cite{MOO70, FUL76}. The radiation flux is thermal for an appropriately chosen accelerated trajectory, and hence an analogy \cite{DAV77,CAR87} can be drawn
with Hawking radiation from a collapsing star \cite{HAW75} that forms a black hole. Interest in this problem has been maintained over the years due to this connection with gravitational physics, but also due to difficulties in obtaining and interpreting results for such systems \cite{OBA01,OBA03,MAR15}. Recently a circuit model approach has been introduced which allows semi-analytical solutions to be obtained for this and related problems \cite{Daiqin2016, Daiqin2017}, allowing clearer exploration of the physics.
\\
\\
Interactions with an accelerated mirror inevitably mixes left and right going modes (in a 1+1 approximation). Hence it has been assumed that the particle production and mixing seen by an inertial observer looking at (say) left moving modes coming from an accelerated mirror are due to loss of information via entanglement to the right going modes. That is, in quantum mechanics it is expected that initial pure states will evolve into pure final states hence, if a mixed state is observed, it is assumed there is some coupling to unobserved parts of the system. By tracing off parts of the system, the observed sub-system may seem mixed due to the information that has been lost. 
\\
\\
However, using the circuit model, it has recently been shown that a single mode squeezed signal sent by a uniformly accelerated observer would, from an operational point of view, be observed to be decohered by an inertial observer \cite{Daiqin2017}. This is in spite of there being no coupling between left and right going modes or to other unobserved degrees of freedom. The key restriction on the observer in this scenario is that they do not possess global information about the modal decomposition of the interaction, but rather are provided with a mode reference from the accelerated source. It is interesting to consider whether similar effects might be present for passive accelerated objects.
\\
\\
In this paper we analyse the effect of the Minkowski vacuum interacting with an accelerated time-delay. The natural modes in the reference frame of an object uniformly accelerating in the right going direction are the right Rindler modes. We model an interaction that delays the right Rindler modes with respect to the left Rindler modes. The delay is passive and doesn't couple left and right going modes. The global effect of such a unitary delay can be analysed straightforwardly in the Schr\"odinger picture and predicts particle production in the Minkowski frame. However the analysis of the statistics expected from particular detection models for inertial Minkowski observers is more complicated.
\begin{figure} [h!]
\centering
\includegraphics[width=0.4\textwidth]{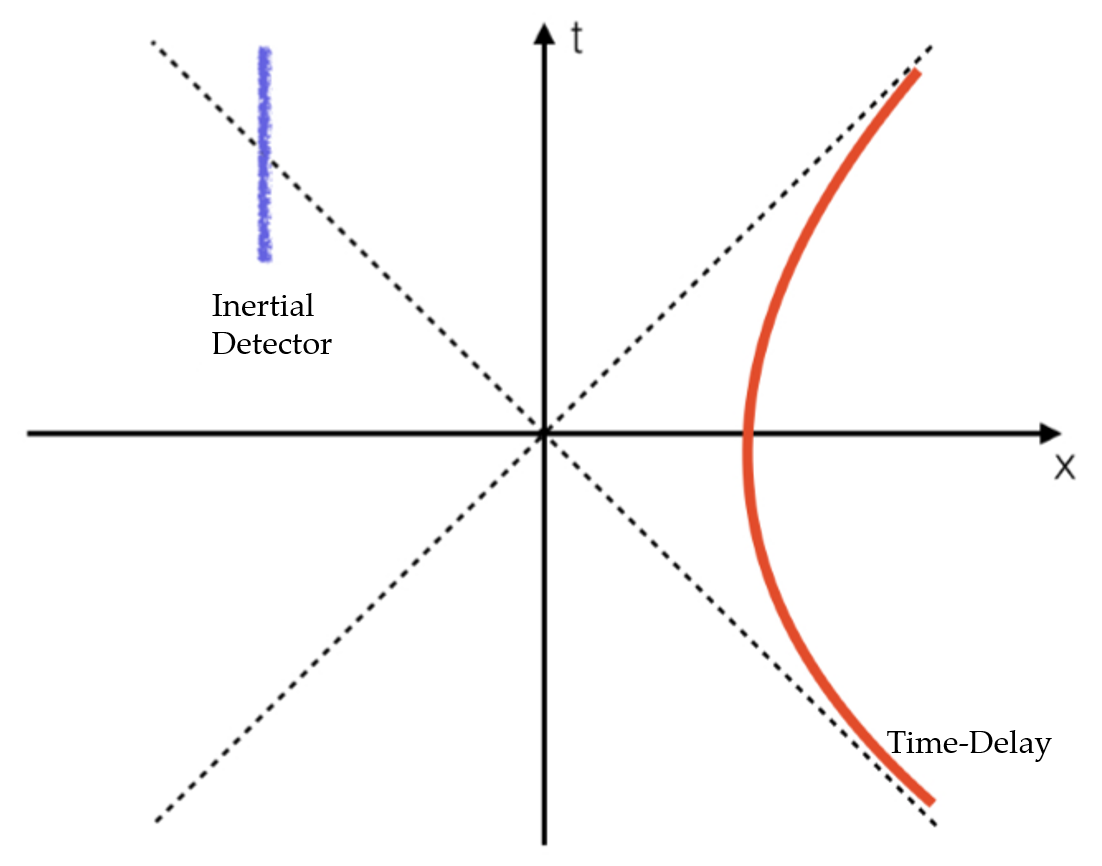}
\caption{The time-delay source moves along the red trajectory. The detector remains stationary along the blue line and holds a broad-band detector.}
\label{fig: Unitary}
\end{figure}
\\
\\
%
%
We adopt the circuit model (input-output) formalism and the self-homodyne detection method for our analysis \cite{Daiqin2017}. In this scheme, the observer's detector is a broad-band ``bucket" detector, looking at all field modes. The mode reference is sent from the accelerated reference frame. The conceptual set-up of such scenario can be found in Fig. 1. This method does not utilize the perturbation method, and given particular conditions, simple, accurate, semi-analytic expression are obtained. This method also has the virtue of analysing the effects of the interaction from an operational point of view which has a strong connection with experimental methodology.
\\
\\
Our paper is set out in the following way. In the following section we introduce our model for the accelerated time delay and derive a global solution in terms of Unruh modes initially in the Minkowski vacuum state. In Section III we introduce the self-homodyne detection model and derive approximate solutions, valid for particular parameter choices. In Section IV we analyse the results and in Section V we make connection with the accelerated mirror under similar conditions and find parameters for which the measurement statistics of the time-delay and mirror coincide. The self-homodyne detection model assumes the mode reference is sent from the source of the interaction in the right Rindler wedge. In Section VI we show that if an additional mode reference is sent from the left Rindler wedge then pure state statistics can be observed for the ``mirror-like" case. Two interesting cases are presented. We discuss and conclude in Section VII.
\section{Accelerated Unitary Time Delay}
\subsubsection{Introducing the Operators}
In this paper we consider a massless scalar bosonic field $\hat{\Phi}$ in (1+1)-dimensional Minkowski space-time. Details on the quantisation method and the definition of the single frequency annihlation/creation opeartors can be found in \cite{Unruh1976, Takagi1986, Crispino2008, Fulling1973}. For simplicity, we only consider the left moving modes in this paper. The single frequency Minkowski annihilation operator is defined as $\hat{e}_k$. It is useful to introduce what is known as the single frequency Unruh operators, $\hat{c}_\omega$ and $\hat{d}_\omega$. The Unruh operators are related to the Minkowski operator in the following way \cite{Daiqin2016, Crispino2008, Birrell1982}:
\begin{align}
&\hat{e}_k=\int d\omega \; A_{k\omega}\hat{c}_{\omega}+B_{k\omega}\hat{d}_{\omega}
\end{align}
	where,
	\begin{equation}
	\begin{aligned}
	A_{k\omega}&=\frac{i\sqrt{2\sinh[\pi\omega/a]}}{2\pi\sqrt{\omega k}}\Gamma[1-i\omega/a]\left(\frac{k}{a}\right)^{i\omega /a} =B_{k\omega}^*
	\end{aligned}
	\end{equation}
Where $\Gamma(x)$ is the gamma function. The Unruh modes are related to the right and left Rindler modes, $\hat{a}_{\omega}$ and $\hat{b}_{\omega}$ respectively, by a two mode squeezing operation. 
	\begin{equation}
	\begin{aligned}
	&\hat{c}_{\omega}= \cosh(r_{\omega}) \hat{a}_{\omega}- \sinh(r_{\omega}) \hat{b}_{\omega}^{\dag} \\
	&\hat{d}_{\omega} 
	= \cosh(r_{\omega}) \hat{b}_{\omega}- \sinh(r_{\omega}) \hat{a}_{\omega}^{\dag} 
	\end{aligned}
	\end{equation}
Where $r_{\omega}\equiv \tanh^{-1}[\exp(-\pi \omega /a)]$ and $a$ is the acceleration of the observer. By inverting equation (3), we obtain the following equations:
	\begin{equation}
	\begin{aligned}
	&\hat{a}_{\omega}= \cosh(r_{\omega}) \hat{c}_{\omega} + \sinh(r_{\omega}) \hat{d}_{\omega}^{\dag} \\
	&\hat{b}_{\omega} 
	= \cosh(r_{\omega}) \hat{d}_{\omega} + \sinh(r_{\omega}) \hat{c}_{\omega}^{\dag} 
	\end{aligned}
	\end{equation}
	These definitions will form the basis of the quantum circuit model (or input-output formalism) which was developed by Su et al. \cite{Daiqin2017, Daiqin2016}. It is noted that we have utilized a different notation to denote the Rindler and Minkowski operators to other authors.
\subsubsection{Introducing the Unitary}
Interactions between uniformly accelerated objects and quantum fields have been studied for many years, however to the best of our knowledge, a time-delay in the Rindler frame has not been previously studied. We first introduce the unitary time-evolution operator in the Rindler frame as follows;
\begin{equation}
\hat{U}_t=e^{-i\hat{H}_R\tau+i\hat{H}_L\overline{\tau}}
\end{equation}
Where $\hat{H}_R$ is the Hamiltonian in the right Rindler wedge and $\hat{H}_L$ is the Hamiltonian in the left Rindler wedge. In the right and left Rindler wedges, the Hamiltonian is defined as follows:
\begin{equation}
\begin{aligned}
\hat{H}_{R}=\int \mathrm{d}\omega \; \omega(\hat{a}_{\omega}^{\dag}\hat{a}_{\omega}), \; \hat{H}_{L}=\int \mathrm{d}\omega \; \omega(\hat{b}_{\omega}^{\dag}\hat{b}_{\omega})
\end{aligned}
\end{equation}
The unitary time delay in the right Rindler wedge can be modelled through the following unitary:
\begin{figure} [h!]
\centering
\includegraphics[width=0.4\textwidth]{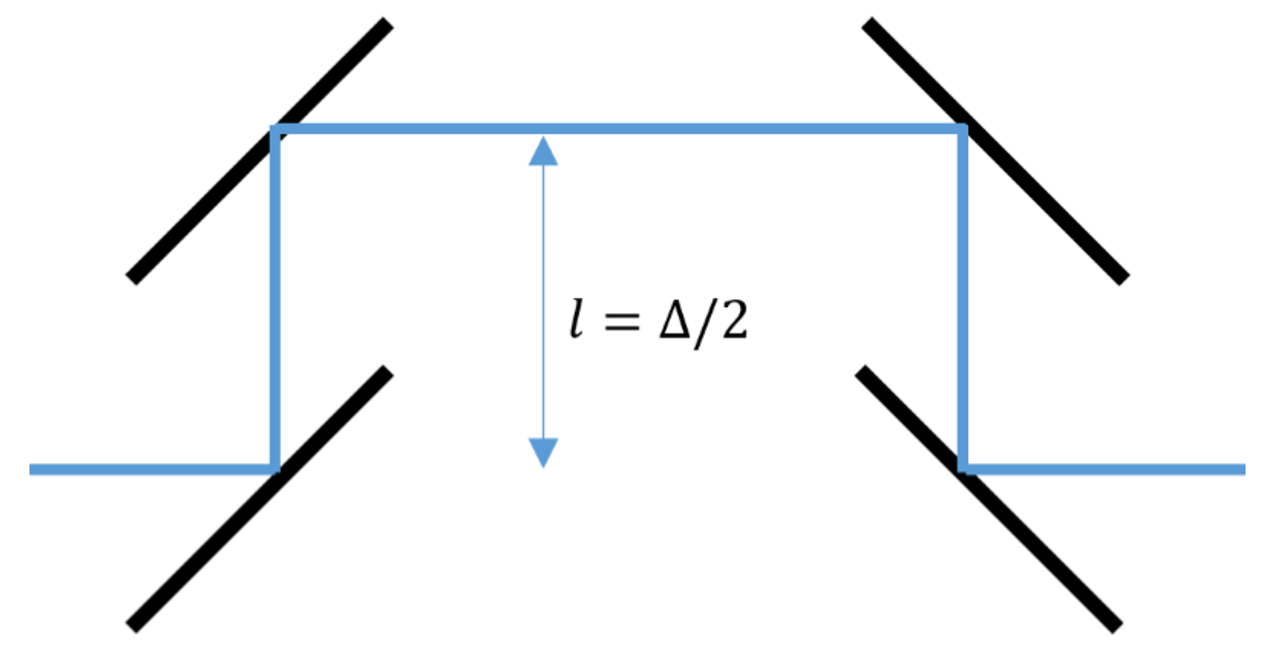}
\caption{The unitary time delay can be modelled through the use of mirrors. It can be seen that the incoming light beam must travel an extra distance of $\Delta$ due to the mirrors. If this mirror arrangement is accelerated, the delay will occur to the Rindler modes in one Rindler wedge.}
\label{fig: Unitary}
\end{figure}
\begin{equation}
\hat{U}=e^{i\hat{H}_{R}\Delta}
\end{equation}
This unitary can be compared with equation (5). It is easy to see that we have set $\tau=-\Delta$ and $\overline{\tau}=0$. We can induce a time delay by accelerating an object which delays the incoming signal by a set time $|\Delta|$. In Fig. 2, a physical example of an accelerated object that could cause such a delay is shown.
\subsubsection{Schr\"odinger Picture}
To have an understanding of what physically occurs to the field, we seek how the Minkowski vacuum evolves under the operator defined in equation (7). The Minkowski vacuum is defined as $\hat{e}_k\ket{0_M}=0, \; \forall k$ while the Rindler vacuum is defined as $\hat{a}_{\omega} \ket{0_R}=\hat{b}_{\omega} \ket{0_R}=0, \; \forall \omega$
We calculate how the Minkowski vacuum transforms under this unitary. First, we look into the output state in terms of the Rindler vacuum (in terms of the accelerated observers) \cite{Crispino2008}:
\begin{widetext}
\begin{equation}
\begin{aligned}
\hat{U} \ket{0_M} =\prod\limits_{\omega}\sqrt{1-\exp[-2\pi \omega/a]} \sum\limits_{n_{\omega}=0}^{\infty} \frac{\exp[-n_{\omega} \pi \omega /a]}{n_{\omega}!} 
(\hat{a}_{\omega}^{\dag} e^{i\Delta\omega} \hat{b}_{\omega}^{\dag})^{n_{\omega}}
\ket{0_R} 
\end{aligned}
\end{equation}
\end{widetext}
We find that we can explicitly write the final state in terms of the Rindler single frequency creation operators from the Rindler vacuum. It is clear that the right and left Rindler observers can measure a pure state by comparing the correlation between the right and left single frequency Rindler particles. 
\\
\\
We now look into the output state in terms of the Minkowski vacuum (in terms of the inertial observers). To do this, we decompose the unitary into a form which can be understood in the Minkowski frame. With lengthy calculations, we can show the following:
\begin{equation}
\hat{U} = \hat{U}_{\hat{d},p}(\omega \Delta)\hat{S}(r,(\theta_2+\theta_{1}))\hat{U}_{\hat{c},p}(\theta_{1})\hat{U}_{\hat{d},p}(\theta_{1})
\end{equation}
Where we have defined the following:
\begin{equation}
\begin{aligned}
r(\omega)&=\cosh ^{-1}(|\cosh(r_\omega)^2 e^{-i \omega \Delta}-\sinh(r_\omega)^2|)
\\
e^{i \theta_1}& \equiv\frac{\cosh(r_\omega)^2 e^{-i \omega \Delta}-\sinh(r_\omega)^2}{|\cosh(r_\omega)^2 e^{-i \omega \Delta}-\sinh(r_\omega)^2|}
\\
e^{i \theta_2}& \equiv\frac{(e^{-i \omega \Delta}-1)}{|e^{-i \omega \Delta}-1|}
\end{aligned}
\end{equation}
The unitary transformation is decomposed into a combination of phase shifters and a two mode squeezer. The two mode squeezer is defined in the following way:
\begin{equation}
\begin{aligned}
&\hat{S}(r,\theta) \equiv \exp [\int \mathrm{d}\omega \; {r(\omega)}(\hat{d}_{\omega}^{\dag} \hat{c}_{\omega}^{\dag}e^{i \theta}-e^{-i \theta}\hat{c}_{\omega} \hat{d}_{\omega})]
\\
&\hat{S}(r,\theta)^{\dag}\hat{c}_{\omega}\hat{S}(r,\theta)=\cosh(r(\omega))\hat{c}_{\omega}+e^{i\theta}\sinh(r(\omega))\hat{d}_{\omega}^{\dag}
\\
&\hat{S}(r,\theta)^{\dag}\hat{d}_{\omega}\hat{S}(r,\theta)=\cosh(r(\omega))\hat{d}_{\omega}+e^{i\theta}\sinh(r(\omega))\hat{c}_{\omega}^{\dag}
\end{aligned}
\end{equation}
Phase shifters are defined in the following way:
\begin{equation}
\begin{aligned}
&U_{\hat{o},p}(\theta) \equiv \exp\left(i\int \mathrm{d}\omega \;\theta(\omega) \hat{o}^{\dag}_{\omega}\hat{o}_{\omega}\right)
\\
&U_{\hat{o},p}(\theta)^{\dag}\hat{o}_{\omega}U_{\hat{o},p}(\theta)=e^{i \theta(\omega) }\hat{o}_{\omega}
\end{aligned}
\end{equation}
By acting this unitary onto the Minkowski vacuum, we find the following:
\begin{equation}
\begin{aligned}
\hat{U} \ket{0_M} &= 
\hat{U}_{\hat{d},p}(\omega \Delta)\hat{S}(r,\theta_2+\theta_{1}) \ket{0_M}
\\
&=\hat{S}(r,(\theta_2+\theta_{1}+\omega\Delta)) \ket{0_M}
\end{aligned}
\end{equation}
As a result, the final state is a pure two-mode squeezed state. Equation (13) makes it clear that different Unruh frequencies are completely uncorrelated from each other, thus we can conclude that the correlations will exist between single frequency Unruh modes. For single frequency, the squeezing strength is $r(\omega)$ and squeezing angle is $\theta_1(\omega)+\theta_2(\omega)+\omega\Delta$. 
\\
\\
From a practical point of view, this information can only be extracted when the observer knows the modal structure of the Unruh modes (which is dependent on the trajectory of the accelerated observer). We impose a key restriction on the inertial observer that they do not possess global information about the modal decomposition of the interaction; the observer has no information on the structure of the incoming signal. That is to say that the signal must be accompanied with information telling the inertial observer where the signal is. 
To address this issue, we implement the self-homodyne detection method. The information of the modal structure will be encoded within the strong coherent signal sent by the source. The inertial observer will only require a broad-band detector and the information of the incoming mode can be analysed via statistical analysis of the particle count.
\section{Self Homodyne Detection on Accelerated Unitary Time Evolution}
\subsection{Self-Homodyne Detection}
We utilize homodyne tomography \cite{Lvovsky2009} to characterise the state of a particular field mode. For Gaussian states, the analysis of the first and second order moment \cite{Weedbrook2012} is sufficient to characterise the Wigner function of a particular output mode \cite{Scully1997}. 
\\
\\
We will utilize self-homodyne detection method to characterise the Wigner function of a particular output mode. This section will introduce the self-homodyne detection method. Self-homodyne detection is conducted through displacing the mode of interest by a large displacement operator $\hat{D}_i(\alpha=|\alpha| e^{i \phi})=\exp[\alpha \hat{o}_i^{\dag}-\alpha^* \hat{o}_i]$. The particle count of such state is compared to the particle count of an output without the presence of the signal for various $\phi$. 
\\
\\
The state with the signal can be created by acting the unitary operator onto the initial state. In the Heisenberg picture, we interpret this as the following:
	\begin{equation}
	\begin{aligned}
	\hat{o}_{i}' & \equiv \hat{U}^{\dag} \hat{o}_{i}\hat{U}
	\end{aligned}
	\end{equation}
The signal that is created is then coupled with a strong coherent signal. In the Heisernberg picture, the operator evolves in the following way:
	\begin{equation}
	\begin{aligned}
	\hat{o}_{i}'' & \equiv \hat{U}^{\dag} \hat{D}^{\dag}_i(\alpha)\hat{o}_{i}\hat{D}_i(\alpha)\hat{U}
	\end{aligned}
	\end{equation}
The photon number operator can be written in the following way:
	\begin{equation}
	\begin{aligned}
	\hat{N}_{i} & \equiv \hat{o}_{i}''^{\dag}\hat{o}_{i}''
	\\ & = |\alpha|^2+|\alpha|\hat{X}_{i}(\phi)+(\hat{o}_i^{\dag}{}'\hat{o}_i')
	\\ & \approx |\alpha|^2+|\alpha|\hat{X}_i(\phi)
	\\ \hat{N}_{0,i} & \equiv \hat{D}_i^{\dag}(\alpha)\hat{o}_{i}^{\dag}\hat{o}_{i}\hat{D}_i(\alpha)
\\	
	&=|\alpha|^2+|\alpha|\hat{X}_{0,i}(\phi)+(\hat{o}_{i}^{\dag}\hat{o}_{i})
	\\ & \approx |\alpha|^2
	\end{aligned}
	\end{equation}
Where we have defined the following:
	\begin{equation}
	\begin{aligned}
\hat{X}_{i}& \equiv \hat{o}_i'e^{-i\phi}+\hat{o}_i^{\dag}{}'
e^{i\phi}
\\ 
\hat{X}_{0,i}& \equiv \hat{o}_ie^{-i\phi}+\hat{o}_i^{\dag}
e^{i\phi}
	\end{aligned}
	\end{equation}
	\\
The approximation in equation (16) is valid when we set $|\alpha|^2 \gg \braket{\hat{o}_i^{\dag}{}'\hat{o}_i'}$ and assume $\braket{\hat{X}_{0,i}(\phi)} \ll |\alpha|$. Utilizing equation (16), we find the following:
	\begin{equation}
	\begin{aligned}
{X}_i(\phi) & \equiv \frac{\braket{\hat{N}_{i}- \hat{N}_{0,i}}}{\sqrt{\braket{\hat{N}_{0,i}}}} \approx  \braket{\hat{X}_{i}(\phi)}	
	\end{aligned}
	\end{equation}
In future calculations, we will differentiate between $\braket{\hat{X}_i(\phi)}$ and $X_i(\phi)$. The first being the explicit expectation value of $\hat{X}_i(\phi)$, while the latter is the approximate value we find via self-homodyne detection method. We take the variance of the expression above and find the following:
	\begin{equation}
	\begin{aligned}
	{V}_i(\phi)& \equiv \frac{(\braket{\Delta \hat{N}_{i}})^2}{{\braket{\hat{N}_{0,i}}}} \approx\braket{\hat{V}_{i}(\phi)}
\end{aligned}
\end{equation}
In cases where the observer does not know the mode in which the signal is sent, the observer is restricted to conducting a measurement over all basis. The mode of interest is amplified by the large coherent signal, thus we can utilize the following assumption:
	\begin{equation}
	\begin{aligned}
	\frac{\hat{N}_{j \neq i}-\hat{N}_{0,j\neq i}}{\braket{\hat{N}_{0,i}}} \approx 0
	\end{aligned}
	\end{equation}
	Utilizing this result and the fact that the total number operator does not change with the change of basis, we conclude the following:
	\begin{equation}
	\begin{aligned}
	(\hat{N}_{i} - \hat{N}_{0,i}) & \approx \sum_j \hat{N}_{j} - \hat{N}_{0,j}  
	\\ & = \sum_n \hat{N}_{n} - \hat{N}_{0,n}  
	\end{aligned}
	\end{equation}
In the equation above, n denotes the complete orthonormal basis set in which the particle number is counted in. By utilizing this approximation, we find the following:
\begin{equation}
\begin{aligned}
 X(\phi) &\equiv \left(\braket{\hat{N}}-\braket{\hat{N_0}}\right)/{\sqrt{\braket{\hat{N}_0}}} \approx X_i(\phi)
\\
V(\phi) &\equiv \left(\braket{\hat{N}^2}-\braket{\hat{N}}^2\right)/\braket{\hat{N}_0}\approx  V_i(\phi) 
\end{aligned}
\end{equation}
Equations (18) and (19) is the homodyne detection measured over a single basis. Equation (22) is a self homodyne detection method which can be implemented when the basis is not well-defined. 
\\
\\
Our observer will be placed in Minkowski space time, and hence we have defined the following:
\begin{equation}
\begin{aligned}
\hat{N}& \equiv \int \mathrm{d}k \; \hat{e}_k^{\dag}{}''\hat{e}_k''
\\ &=\int \mathrm{d}\omega \; \hat{c}_{\omega}^{\dag}{}''\hat{c}_{\omega}''+\hat{d}_{\omega}^{\dag}{}''\hat{d}_{\omega}''
\\
\hat{N}_0 & =\; \hat{D}_i (\alpha)^{\dag} \left(\int \mathrm{d}\omega \; \hat{c}_{\omega}^{\dag}{}\hat{c}_{\omega}+\hat{d}_{\omega}^{\dag}\hat{d}_{\omega}\right) \hat{D}_i (\alpha)
\end{aligned}
\end{equation}
Equations (22) and (23) will form the foundation of the self-homodyne detection method. We now analyse how these equations can be put into a more useful form. To do this, we define the following operators:
\begin{equation}
\begin{aligned}
\hat{c}_{\omega }'' \equiv& \hat{U}^{\dag} \hat{D}_i (\alpha)^{\dag} \hat{c}_{\omega} \hat{D}_i (\alpha)\hat{U}
\\
\hat{d}_{\omega }'' \equiv& \hat{U}^{\dag}\hat{D}_i (\alpha)^{\dag} \hat{d}_{\omega} \hat{D}_i (\alpha)\hat{U} 
\\
\hat{c}_{\omega }' \equiv& \hat{U}^{\dag} \hat{c}_{\omega} \hat{U} 
\\
\hat{d}_{\omega }' \equiv& \hat{U}^{\dag}  \hat{d}_{\omega}' \hat{U} 
\end{aligned}
\end{equation}
We also define the following values:
\begin{equation}
\begin{aligned}
g_c(\omega) \equiv & \;  \hat{c}_\omega{}''-\hat{c}_\omega{}'
\\
g_d(\omega) \equiv & \; \hat{d}_\omega{}''-\hat{d}_\omega{}'
\end{aligned}
\end{equation}
$g_c(\omega)$ and $g_d(\omega)$ are generally in the order of magnitude of $\alpha$. Utilizing these expressions, and the fact that $\hat{c}_{\omega}\ket{0_M}=\hat{d}_{\omega}\ket{0_M}=0$, we find that the quadrature amplitude can be calculated as follows:
\begin{equation}
\begin{aligned}
 X(\phi) \approx & \frac{2 Re[ \int\mathrm{d}\omega\; g_c(\omega)^*\braket{\hat{c}_{\omega }'}+g_d(\omega)^*\braket{\hat{d}_{\omega } '}]}{[\int\mathrm{d}\omega\; |g_c(\omega)|^2+|g_d(\omega)|^2]^{1/2}}
\end{aligned}
\end{equation}
We have neglected all values which are in the order of $1/{|\alpha|}$, as we can set $\alpha$ to be arbitrarily large. We now proceed onto calculating the quadrature variance. To do this, we first simplify the following expectation values:
\begin{widetext}
\begin{equation}
\begin{aligned}
 \braket{\hat{c}_{\omega}^{\dag}{}''\hat{c}_{\omega}{}'' \hat{c}_{\omega'}^{\dag}{}''\hat{c}_{\omega'}{}''} - \braket{\hat{c}_{\omega}^{\dag}{}''\hat{c}_{\omega}{}''}\braket{ \hat{c}_{\omega'}^{\dag}{}''\hat{c}_{\omega'}{}''}\approx \; 
 & g_c(\omega)g_c(\omega')^*\braket{ \delta(\omega-\omega')+ 2\hat{c}_{\omega}^{\dag}{}'\hat{c}_{\omega'}{}'}+2 Re[g_c(\omega)g_c(\omega')\braket{ \hat{c}_{\omega}^{\dag}{}'\hat{c}_{\omega'}^{\dag}{}'}]
\\
 \braket{\hat{d}_{\omega}^{\dag}{}''\hat{d}_{\omega}{}'' \hat{d}_{\omega'}^{\dag}{}''\hat{d}_{\omega'}{}''} - \braket{\hat{d}_{\omega}^{\dag}{}''\hat{d}_{\omega}{}''}\braket{ \hat{d}_{\omega'}^{\dag}{}''\hat{d}_{\omega'}{}''} \approx  \;
 & g_d(\omega)g_d(\omega')^*\braket{ \delta(\omega-\omega')+2 \hat{d}_{\omega}^{\dag}{}'\hat{d}_{\omega'}{}'}+2 Re[g_d(\omega)^*g_d(\omega')^*\braket{ \hat{d}_{\omega}{}'\hat{d}_{\omega'}{}'}]
\\
 \braket{\hat{c}_{\omega}^{\dag}{}''\hat{c}_{\omega}
{}'' \hat{d}_{\omega'}^{\dag}{}''\hat{d}_{\omega'}{}''}- \braket{\hat{c}_{\omega}^{\dag}{}''\hat{c}_{\omega}{}''}\braket{ \hat{d}_{\omega'}^{\dag}{}''\hat{d}_{\omega'}{}}
 \approx  \; &  2 Re[ g_c(\omega)^*g_d(\omega')^*\braket{\hat{c}_{\omega}{}'\hat{d}_{\omega'}{}'}+g_c(\omega)g_d(\omega')^*\braket{ \hat{c}_{\omega}^{\dag}{}'\hat{d}_{\omega'}{}'}]
\end{aligned}
\end{equation}
\end{widetext}
We have neglected the terms which are 0th and 1st order in $|\alpha|$. The remaining terms are 2nd order in $|\alpha|$.
We now simplify this expression a little more by assuming that the displacement operator is applied onto the right Rindler mode; $\hat{D}_g(\alpha=|\alpha| e^{i\phi})$. The explicit expression of the displacement operator is defined in equation (32) and the explicit expressions for equation (25) are calculated in equation (34).
 Utilizing the expression found in equation (34), it is easy to show that $g_c(\omega)g_c(\omega')^*$, $g_d(\omega)g_d(\omega')^*$ and $g_c(\omega)^*g_d(\omega')^*$ are not dependent on $\phi$. We can also show that $g_c(\omega)g_c(\omega')$, $g_d(\omega)^*g_d(\omega')^*$ and $g_c(\omega)g_d(\omega')^*$ are proportional to $e^{2i \phi}$. Thus, we can split the variance into the part that is phase insensitive and that which is phase sensitive:
	\begin{equation}
	\begin{aligned}
	V(\phi)\approx 1+ V_1+V_2(\phi) 
	\end{aligned}
	\end{equation}
Where we have defined the following:
\begin{widetext}
\begin{equation}
\begin{aligned}
V_1 &\equiv\frac{\int \mathrm{d}\omega \mathrm{d}\omega' \;
2\{g_c(\omega)g_c(\omega')^*\braket{\hat{c}_{\omega}^{\dag}{}'\hat{c}_{\omega'}{}'}
+g_d(\omega)g_d(\omega')^* \braket{\hat{d}_{\omega}^{\dag}{}'\hat{d}_{\omega'}{}'}
+2Re[g_c(\omega)^*g_d(\omega')^* \braket{\hat{c}_{\omega}{}'\hat{d}_{\omega'}{}'}]\}}{\int \mathrm{d}\omega \; |g_c(\omega)|^2+|g_d(\omega)|^2}
\\
V_2(\phi) &\equiv \frac{2\{Re[g_c(\omega)g_c(\omega')\braket{ \hat{c}_{\omega}^{\dag}{}'\hat{c}_{\omega'}^{\dag}{}'}+g_d(\omega)^*g_d(\omega')^*\braket{ \hat{d}_{\omega}{}'\hat{d}_{\omega'}{}'} +2g_c(\omega)g_d(\omega')^*\braket{ \hat{c}_{\omega}^{\dag}{}'\hat{d}_{\omega'}{}'}]\}}{\int \mathrm{d}\omega \; |g_c(\omega)|^2+|g_d(\omega)|^2}
\end{aligned}
\end{equation}
Alternatively, we can write $V_2(\phi)$ in the following way:
\begin{equation}
\begin{aligned}
V_2(\phi) &= - \overline{V}_{2} \times \cos(2\phi-\theta)
\\	
\overline{V}_2 & = \left| 
\frac{2[g_c(\omega)g_c(\omega)\braket{ \hat{c}_{\omega}^{\dag}{}'\hat{c}_{\omega'}^{\dag}{}'}+g_d(\omega)^*g_d(\omega)^*\braket{ \hat{d}_{\omega}{}'\hat{d}_{\omega'}{}'} +2g_c(\omega)g_d(\omega)^*\braket{ \hat{c}_{\omega}^{\dag}{}'\hat{d}_{\omega'}{}'}]}{\int \mathrm{d}\omega \; |g_c(\omega)|^2+|g_d(\omega)|^2}
\right|
\\
\theta &= -\arccos\left(-\frac{V_2(0)}{\overline{V}_2}\right) 
\end{aligned}
\end{equation}
\end{widetext}
Equation (30) decomposes $V_2(\phi)$ into two different parts. $\overline{V}_2$ is a measure of how large the squeezing effect is, and $\theta$ is the angle in which the state is squeezed. We now have all the necessary tools to calculate the statistics of a certain mode via self-homodyne detection method. In the next section, we will introduce the circuit model to calculate the correlation function that are of interest.
\subsection{Circuit Model}
In this section, we implement the circuit model to calculate the first and second order mode moment. The unitary only interacts with the right Rindler mode, thus we expect no radiation in the left Rindler wedge from the accelerated time-delay. We can analyse the Wigner function of the radiation from the accelerated time-delay source by analysing the right Rindler statistics. Hence this section will focus on analysing the statistics of an arbitrary right Rindler mode $\hat{a}_g$. This Rindler mode is defined in the following way:
\begin{equation}
\begin{aligned}
\hat{a}_g &\equiv \int \mathrm{d} \omega \; \hat{a}_{\omega}g(\omega)
\end{aligned}
\end{equation}
Where $g(\omega)$ is an arbitrary normalised positive frequency mode. Thus we introduce the displacement operator in the following way: 
\begin{equation}
\hat{D}_g(\alpha=|\alpha|e^{i\phi})\equiv \exp(\alpha\hat{a}_g^{\dag}-\alpha^*\hat{a}_g)
\end{equation}
Any arbitrary bosonic operator can be written as a superposition of the part that overlaps with $\hat{a}_g$ and a part which is orthonormal to $\hat{a}_g$ \cite{Daiqin2017, Rohde2007}:
\begin{equation}
\begin{aligned}
\hat{o}=(\hat{o}-([\hat{o},\hat{a}_g^{\dag}]\hat{a}_g+[\hat{a}_g,\hat{o}]\hat{a}_g^{\dag}))+([\hat{o},\hat{a}_g^{\dag}]\hat{a}_g+[\hat{a}_g,\hat{o}]\hat{a}_g^{\dag}{})
\end{aligned}
\end{equation}
We have decomposed an arbitary bosonic operator $\hat{o}$ into two terms; the second term in the braket is affected by a unitary that acts on a particular mode $\hat{a}_g$, and the first term in the braket remains unaffected. We now have the necessary tools to introduce the input-output relations \cite{Daiqin2017, Obadia2001, Daiqin2016}. We expand equation (25) utilizing this decomposition to find:
\begin{equation}
\begin{aligned}
g_c(\omega)&= g(\omega)^*\cosh(r_\omega)\alpha
\\
g_d(\omega)&=- g(\omega)\sinh(r_\omega)\alpha^*
\end{aligned}
\end{equation}
Likewise, we can calculate how the Unruh operators evolve by utilizng equations (3) and (4):
\\
\\
\begin{widetext}
\begin{equation}
\begin{aligned}
 \hat{a}_{\omega}' &=\hat{a}_{\omega}e^{-i\omega \Delta} 
\\
\hat{c}_{\omega}' &=\hat{c}_{\omega}+\cosh(r_\omega)(e^{-i \omega \Delta}-1)\hat{a}_{\omega}=\hat{c}_{\omega}\left(\cosh(r_{\omega})^2 e^{-i \omega \Delta}-\sinh(r_{\omega})^2\right)+\hat{d}_{\omega}^{\dag} \cosh(r_{\omega})\sinh(r_{\omega})(e^{-i\omega \Delta}-1)
\\
\hat{d}_{\omega}' &=\hat{d}_{\omega}+\sinh(r_\omega)(e^{i \omega \Delta}-1)\hat{a}_{\omega}^{\dag}=\hat{d}_{\omega} (\cosh(r_{\omega})^2-\sinh(r_{\omega})^2 e^{i \omega \Delta})+ \hat{c}_{\omega}^{\dag} \cosh(r_{\omega})\sinh(r_{\omega})(e^{i \omega \Delta}-1)
\end{aligned}
\end{equation}
\end{widetext}
We can calculate the quadrature variance and amplitude by utilizing equations (22), (34) and (35).
\subsection{Quadrature Amplitude and Variance}
By utilizing the fact that $\braket{\hat{c}_{\omega}'}=0$ and $\braket{\hat{d}_{\omega}'}=0$, we find the following:
\begin{equation}
\begin{aligned}
X(\phi)=0
\end{aligned}
\end{equation}
We note there are some complications to equations (36) and (37) which will be addressed in the appendix  Sec. \ref{sec: Reference}. We utilize the formalism we introduced in equation (28) to (30) to calculate the quadrature variance. Utilizing the correlation functions that are calculated in the appendix Sec. VIII, we find the following:
\begin{widetext}
\begin{equation}
\begin{aligned}
V(\phi) & \approx 1+ \frac{8 \int \mathrm{d}\omega \; (1+2 \sinh(r_{\omega})^2) \cosh(r_{\omega})^2 \sinh(r_{\omega})^2 |g(\omega)|^2(1-\cos (\omega \Delta))}{(1+2 I_s)}
\end{aligned}
\end{equation}
\end{widetext}
\begin{equation}
\begin{aligned}
\\I_c &\equiv \int \mathrm{d}\omega \; \cosh(r_{\omega})^2 |g(\omega)|^2
\\
I_s &\equiv \int \mathrm{d}\omega \; \sinh(r_{\omega})^2 |g(\omega)|^2
\end{aligned}
\end{equation}
As a result, $V_1=V(\phi)-1$ and $V_2=0$. We were able to obtain a completely general simple semi-analytic expression for the variance through the use of the input-output formalism. Furthermore, this expression can be obtained in an actual experiment through analysing the photon count statistics. This expression will be analysed numerically in order to gain further understanding of the statistics of the signal.
\section{Statistical Analysis of Accelerated Unitary Time Evolution via Self Homodyne Detection}
\subsection{Self Homodyne Measurement in Rindler Vacuum}
Before examining the statistics of the signal created by a unitary time-delay, we analyse the statistics of the Minkowski vacuum in the right Rindler frame; a thermal statistics.  
We examine the statistics of the right Rindler frame via self-homodyne detection. As usual, we introduce a large reference signal by the displacement operator $\hat{D}_g(\alpha)$. We are interested in the following scenario:
\begin{equation}
\begin{aligned}
\hat{U}&=1
\\
\hat{N}_{R}&=\int \mathrm{d}\omega \; \hat{a}_\omega^{\dag}{}''\hat{a}_\omega''
\\
\hat{N}_{0,R}&=\hat{D}_g(\alpha)^{\dag}\left(\int \mathrm{d}\omega \; \hat{a}_\omega^{\dag}\hat{a}_\omega\right)\hat{D}_g(\alpha)
\end{aligned}
\end{equation}
Following similar steps to the previous sections, we calculate the following:
\begin{equation}
\begin{aligned}
\braket{\hat{a}_{\omega}'}&=0
\\
\braket{\hat{a}_{\omega}'\hat{a}_{\omega}'}&=0
\\
\braket{\hat{a}_{\omega}'^{\dag}\hat{a}_{\omega}'}&=\sinh(r_\omega)^2
\\
g_a(\omega)&=g^*(\omega)\alpha
\end{aligned}
\end{equation}
As a result, the quadrature amplitude and variance are calculated to be as follows:
\begin{equation}
\begin{aligned}
X_{vac}(\phi)=0
\\
V_{vac}(\phi)=1+2I_s
\end{aligned}
\end{equation}
This describes the statistics of a thermal bath, and it will be compared with the result obtained in equation (37).

\subsection{Numerical Analysis of Variance}
The analysis with the Schr\"odinger picture showed that there are no correlations between different Rindler/Unruh frequency modes. As a result, we are interested in the right Rindler single frequency statistics due to the unitary. As it is difficult to consider a normalized single frequency mode, we consider a localised Gaussian wave-packet mode in the right Rindler frame:
\begin{figure}[h!]
\centering
\includegraphics[width=0.48\textwidth]{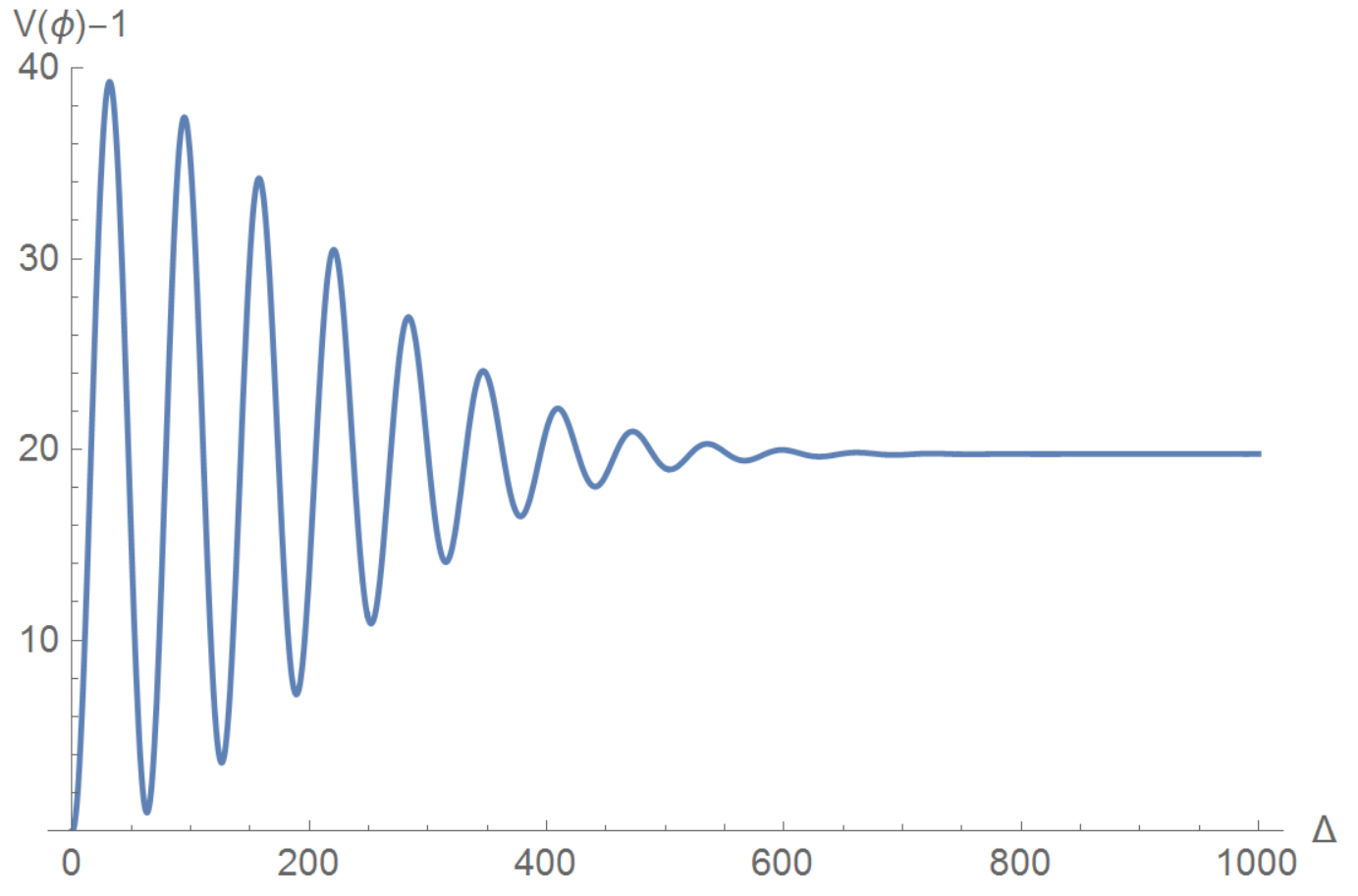}
\caption{The graph above demonstrates how $\Delta$ affects the variance of the signal detected by the Minkowski detector. We have utilized the following settings: $a=1$, $\omega_0=0.1$ and $\delta=0.005$.}
\end{figure}
\begin{equation}
g(\omega;\omega_0, \delta, v_c) \equiv B\sqrt{\omega}(\frac{1}{2\pi \delta^2})^{1/4} \exp\left[-\frac{(\omega-\omega_0)^2}{4\delta^2}-i\omega v_c\right] 
\end{equation}
where $B$ is the normalization constant, $\omega_0$ is the central frequency, $\delta$ is the bandwidth of the wavepacket mode and $v_c$ is the central position of the Gaussian wavepacket mode. By restricting ourselves to $\delta<0.4 \omega_0$, the approximation $B \approx 1/\sqrt{\omega_0}$ is valid. In this section we compute $V_1$ which characterises the deviation of the variance from the shot noise. We first analyse how $\Delta$ affects the variance. 
\\
\\
\noindent We find that the variance remains roughly constant for $\Delta > \delta^{-1}$ from Fig. 3. This is the regime where the overlap between the delayed mode is roughly zero; $[\hat{a}_g{}',\hat{a}_g^{\dag}] \approx 0$. While the original and delayed modes are overlapping, we observe sinusoidal waves. This can be understood as a result of the wave-packet modes becoming correlated/anti-correlated. When the two modes are out of phase by $\pi$, we observe a local maxima in the variance. The local minima corresponds to when the two modes are in phase with each other. The amplitude of the sinusoidal waves decreases due to the decrease in the overlap between the two modes.
\begin{figure}[h!]
\centering
\includegraphics[width=0.48\textwidth]{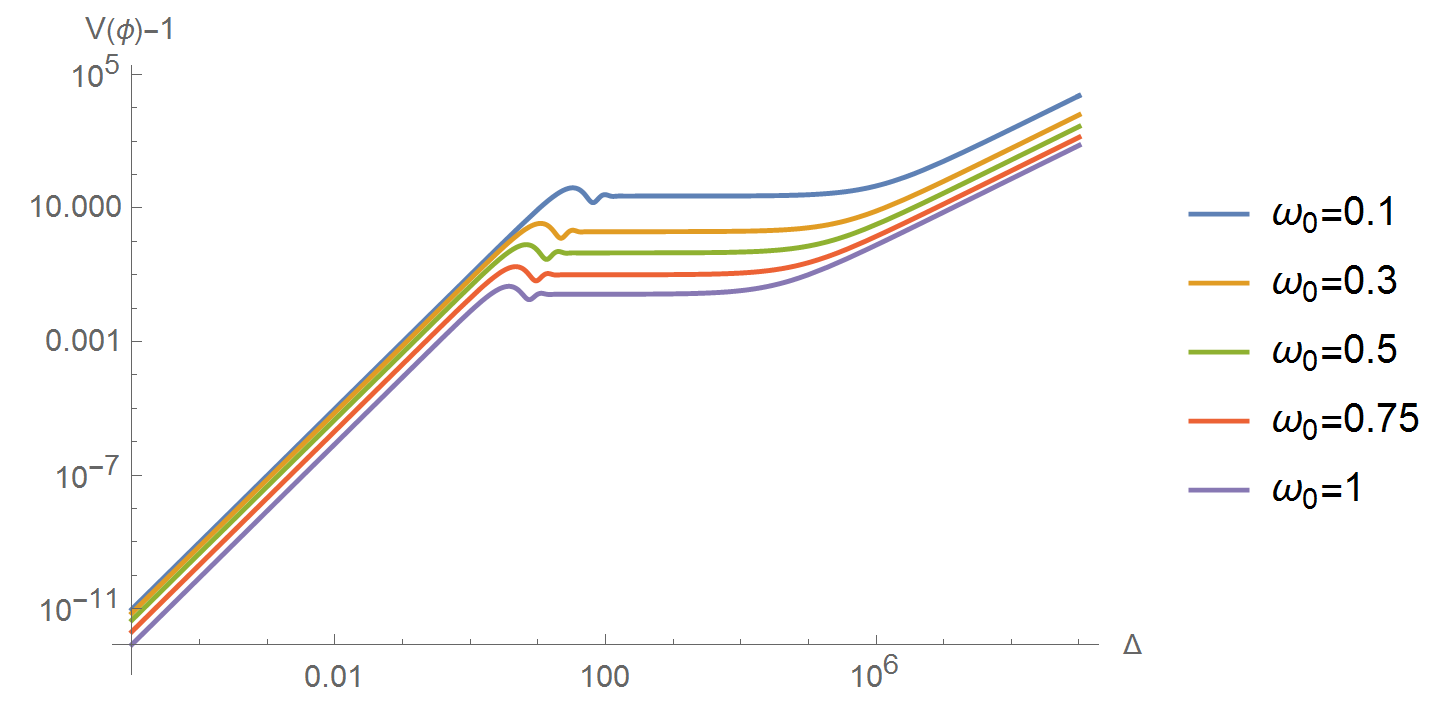}
\caption{The graph above demonstrates how $\Delta$ affects the variance of the signal detected by the Minkowski detector. We have utilized the following settings: $a=1$ and $\delta=0.2\omega_0$.}
\end{figure}
\\
\\
Fig. 4 analyses how the variance increases for various $\omega_0$. We find that the variance increases quadratically for small $\Delta$. This regime corresponds to when the two modes still overlap with each other. As Fig 4. has a significantly larger value of $\delta$ than Fig. 3, we only see one oscillation before the variance becomes constant. The regime where the variance is constant can be interpreted as the regime when the two modes no longer overlapping with each other.
\\
\\
The variance starts to increase again for a sufficiently large $\Delta$. With further analysis, it can be shown that the variance increases due to low frequency contributions around $\omega=0$. As a result this regime can be interpreted as an artifact of assuming the delay also applies to ultra low frequencies. As this is physically unlikely, we are not interested in this regime. The low frequency contributions can be suppressed by setting $\delta \ll \omega_0$. 
\\
\\
We can obtain numerical results that are similar to the single frequency statistics by considering the regime where the variance is roughly constant. This can be done by setting $\delta \ll \omega_{0}$ and setting $1/\delta \ll \Delta$. Fig. 5 is a plot which demonstrates the statistics of the signal; how $\omega_0$ affects the variance.
\begin{figure} [h!]
\centering
\includegraphics[width=0.45\textwidth]{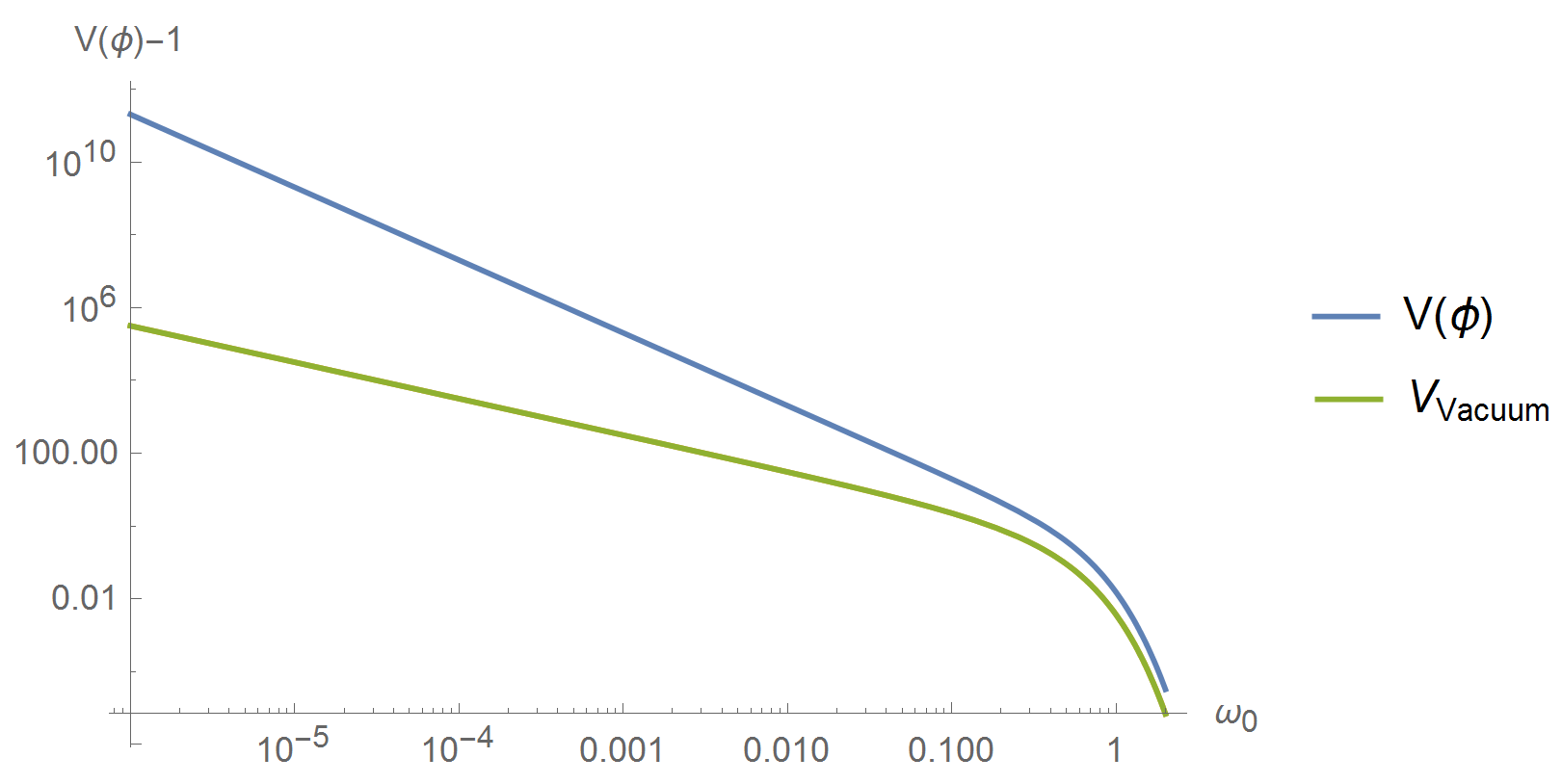}
\caption{The graph above is a plot of which demonstrates the statistics of the state. We have utilized the following settings: $\Delta=10^8, \delta=0.05 \omega_0$.}
\label{fig: Unitary}
\end{figure}
\\
\\
Fig. 5 compares the variance of $V_{vac}$ (Eq. 41) and $V(\phi)$. 
For low $\omega_0$, it is found that $V_{vac}$ is inversely proportional to $\omega_0$, while $V(\phi)$ is inversely proportional to $\omega_0 ^2$. This demonstrates the statistics of a thermal state and the state created by an accelerated time delay on Minkowski vacuum are quite different in general. On the other hand, it is interesting to note that for high $\omega_0$ the characteristics of the variances are quite similar. This figure can be summarised with the following equations:
\begin{equation}
\begin{aligned}
V(\phi,\omega) &= 1+ 2\csch(\pi \omega/a)^2
\\
V_{vac}(\phi,\omega) &= 1+ 2\sinh(r_{\omega})^2
\end{aligned}
\end{equation}
The output state has fluctuation above the shot noise, and hence seems mixed. Through the analysis in the Schr\"odinger picture, we know that the output state is a two mode squeezed state. However, this does not mean that the local observer can easily observe a two-mode squeezed state. The analysis via self homodyne detection demonstrates that when the local observer only looks at the statistic of the radiation from the accelerated time-delay source, the radiation seems noisy. As a result, the output state has apparent decoherence.
\\
\\
This apparent decoherece can be traced back to the underlying vacuum correlations that existed between the right and left Rindler modes. The unitary distorted this correlation. We conjecture that the inertial observer can observe a final pure state when this correlation is extracted. Due to technical reasons, extracting this correlation for a time-delay is difficult. As a result, we consider another passive unitary, where some of the correlations can be extracted more easily.
\section{Mirror}
In this section we briefly consider another passive unitary; an accelerated mirror. This unitary has been considered in many literature in the past \cite{Crispino2008}. The Minkowski frequency statistics of the outcome has been considered by Su et al \cite{Daiqin2016}. In this paper we consider the statistics with respect to Rindler frequencies.
\\
\\
By introducing the right and left moving modes, $\hat{a}_{\omega,1}$ and $\hat{a}_{\omega,2}$ respectively, we can introduce the mirror operator as follows:
\begin{equation}
\hat{U}_M \equiv \exp [\int \mathrm{d}\omega\; \theta_\omega(\hat{a}_{\omega,1}^{\dag}\hat{a}_{\omega,2}-\hat{a}_{\omega,2}^{\dag}\hat{a}_{\omega,1})]
\end{equation}
The operators evolve under this unitary in the following way:
\begin{equation}
\begin{aligned}
\hat{a}_{\omega,1}' &= \hat{a}_{\omega,1}\cos (\theta_\omega)+\hat{a}_{\omega,2}\sin (\theta_\omega)
\\
\hat{a}_{\omega,2}' &= \hat{a}_{\omega,2}\cos (\theta_\omega)-\hat{a}_{\omega,1}\sin (\theta_\omega)
\end{aligned}
\end{equation}
Following the same method taken before, we find that the Unruh operators evolve in the following way:
\begin{equation}
\begin{aligned}
\hat{c}_{\omega}' &= \hat{c}_{\omega}+\cosh(r_\omega)(\hat{a}_{\omega,1}(\cos (\theta_\omega)-1)+\hat{a}_{\omega,2}\sin (\theta_\omega))
\\
\hat{d}_{\omega}' &= \hat{d}_{\omega}-\sinh(r_\omega)(\hat{a}_{\omega,1}^{\dag}(\cos (\theta_\omega)-1)+\hat{a}_{\omega,2}^{\dag}\sin (\theta_\omega))
\end{aligned}
\end{equation}
Utilizing this result, we find the following:
\begin{equation}
\begin{aligned}
\braket{\hat{c}_{\omega}'} &=0
\\
\braket{\hat{d}_{\omega}'} &=0
\end{aligned}
\end{equation}
Hence, the quadrature variance is $X_M(\phi)=0$. We now proceed onto calculating the variance. To do this, we first calculate the following correlation functions:
\begin{equation}
\begin{aligned}
\braket{0_M|\hat{c}_{\omega}^{\dag}{}'\hat{c}_{\omega'}{}'|0_M}&= F_{M,1}(\omega) \delta(\omega-\omega')
\\
\braket{0_M|\hat{d}_{\omega}^{\dag}{}'\hat{d}_{\omega'}{}'|0_M}&=F_{M,1}(\omega) \delta(\omega-\omega')
\\
\braket{0_M|\hat{c}_{\omega}{}'\hat{c}_{\omega'}{}'|0_M}&=0\\
\braket{0_M|\hat{d}_{\omega}{}'\hat{d}_{\omega'}{}'|0_M}&=0
\\
\braket{0_M|\hat{c}_{\omega}{}'\hat{d}_{\omega'}{}'|0_M}&=F_{M,2}(\omega) \delta(\omega-\omega')
\\
\braket{0_M|\hat{c}_{\omega}^{\dag}{}'\hat{d}_{\omega'}{}'|0_M}&=0
\end{aligned}
\end{equation}
Where we have defined the following:
\begin{equation}
\begin{aligned}
F_{M,1}&\equiv 2\cosh(r_\omega)^2\sinh(r_\omega)^2(1-\cos(\theta))
\\
F_{M,2}&\equiv -\cosh(r_\omega)\sinh(r_\omega)(1+2\sinh(r_\omega)^2)(1-\cos(\theta))
\end{aligned}
\end{equation}
It is interesting to compare this result to the result found in equation (58). We utilize the same procedure as before to calculate the variance. The values for $g_c(\omega)$ and $g_d(\omega)$ are the same as the ones calculated in equation (34). 
\begin{widetext}
\begin{equation}
\begin{aligned}
V_M(\phi) & \approx 1+ \frac{8 \int \mathrm{d}\omega \; (1+2 \sinh(r_{\omega})^2) \cosh(r_{\omega})^2 \sinh(r_{\omega})^2 |g(\omega)|^2(1-\cos (\theta))}{(1+2 I_s)}
\end{aligned}
\end{equation}
\end{widetext} 
We obtain the single frequency statistic by substituting $|g(\omega)|^2$ with a delta-function:
\begin{equation}
V_M(\phi,\omega)=1+2\csch(\pi\omega/a)^2(1-\cos(\theta))
\end{equation}
By comparing this equation to equation (41), we find that the two results coincide when $\theta=\pi/2$. The condition in which the result in equation (43) was valid is when $\Delta \gg 1/\delta$. Both conditions correspond to cases when $[\hat{a}_{\omega}',\hat{a}_{\omega}^{\dag}]=0$, which is when the overlap of the displaced mode and the original mode is 0. This could be understood as a measure of how much the vacuum correlation between the left Rindler and right Rindler mode has been distorted. It is interesting to note that the variance is highest when we set $\theta=\pi$. An analogous result of this can be understood as the local maxima that are observed in Fig. 3.
\section{Purification of the the Output State}
In this section, we introduce several special cases where purification of the state can be observed. We extract the correlation for the mirror case, as $F_{M,2}(\omega)$ is a real valued function. This makes extraction of correlation much simpler compared to $F_2(\omega)$ which is a complex valued function.
\\
\\
The strategy is now not only to place the displacement for self-homodyne on the right Rindler mode, but also to displace the left Rindler mode. We displace the right and left Rindler wedge by the following displacement operators:
\begin{equation}
\begin{aligned}
\hat{D}_{g,R}(\alpha=|\alpha|e^{i \phi})\equiv \exp(\alpha\hat{a}_g^{\dag}-\alpha^{*}\hat{a}_g)  
\\
\hat{D}_{g,L}(\beta=|\beta|e^{-i \phi})\equiv \exp(\beta\hat{b}_g^{\dag}-\beta^{*}\hat{b}_g)
\end{aligned}
\end{equation}
where we have defined the following:
\begin{equation}
\begin{aligned}
\hat{b}_g=\int \mathrm{d}\omega \; g(\omega)^*\hat{b}_\omega 
\end{aligned}
\end{equation}
In this case, we calculate $g_c(\omega)$ and $g_d(\omega)$ to be:
\begin{equation}   
\begin{aligned}    
g_c(\omega)& = g(\omega)e^{i \phi} (\cosh(r_\omega) |\alpha| - \sinh(r_\omega) |\beta|)
\\        
g_d(\omega)& = g(\omega)^*e^{-i \phi}(\cosh(r_\omega) |\beta| - \sinh(r_\omega) |\alpha|)
\end{aligned}        
\end{equation}   
As a result, $g_c(\omega)$ and $g_d(\omega)^*$ are proportional to $e^{i \phi}$. Thus, we can use equations (28), (29), (48) and (54) to calculate the variance of the output state. We find that $V_2=0$. For simplicity, we consider the single frequency limit. When we set either of the following:
\begin{equation}
\begin{aligned}
\frac{|\alpha|}{|\beta|}=\frac{\cosh(r_\omega)^2+\sinh(r_\omega)^2}{2\cosh(r_\omega)\sinh(r_\omega)}
\\
\frac{|\beta|}{|\alpha|}=\frac{\cosh(r_\omega)^2+\sinh(r_\omega)^2}{2\cosh(r_\omega)\sinh(r_\omega)}
\end{aligned}
\end{equation}
We find that the single frequency variance is:
\begin{equation}
V(\omega,\phi)=1
\end{equation}
As a result, we observe vacuum. This means that the left Rindler mode is perfectly anti-correlated with the noisy particles we observed coming from the right Rindler wedge. Furthermore, if we set $|\alpha|=|
\beta|$ we find that the single frequency variance is:
\begin{equation}
V(\omega,\phi)= 1+2\cosh(r_{\omega})\sinh(r_{\omega})(\sinh(2r_\omega)-\cosh(2r_\omega))
\end{equation}
The noise is less than vacuum. As a result, the output state has a characteristic of a two-mode squeezed state. This suggests there is entanglement between the particles that is coupled with the right and left Rindler frequencies, as observed by the Minkowski observer. By fully characterising this correlation, we would be able purify the output state.
\\
\\
This section highlighted some simple scenarios where purification of the state could found, implying the presence of entanglement. Full characterisation of this entanglement is not a simple task, and exceeds the scope of this paper.
\section{Conclusion}
In this paper we looked into the effect of accelerated unitary time-delay. Through the Schr\"odinger picture, we showed that the output state is a two-mode squeezed state. We continued onto analysing what an inertial observer would observe due to this unitary via self-homodyne detection. We showed that the radiation from the accelerated time-delay source would be observed to be noisy according to an inertial observer. As a result, accelerated time-delay causes an apparent decoherence. We propose that the information is hidden in the vacuum noise that existed in the left Rindler wedge. We conducted some further research into the mirror case, and we showed that indeed correlations existed in the left Rindler wedge. 
\\
\\
We believe that, from an operational point of view, the extraction of this information is not practical. The stationary observer wants to extract information out of the signal that is sent. As a result, they would not know the source of the signal. The stationary observer only has access to the physical signal (radiation) that is created from the right Rindler observer. The sender is causally disconnected from the left Rindler wedge, and the sender cannot tell the observer which mode the vacuum correlations would be hidden in. As the stationary observer has no clue where the vacuum correlation is hidden (i.e. only has access to the decohered physical signal), according to the stationary observer the system has apparently decohered. 
\\
\\
The only method in which a pure state can be observed by the stationary observer is if we considered a scenario where two parties agreed on which signal is sent by the right Rindler observer. They calculate where the vacuum correlations will be hidden due to that particular signal. The two parties then follow the left/right Rindler trajectories. The sender in the right Rindler wedge sends a signal, and the other party in the left Rindler tells the stationary observer where the vacuum correlation would be hidden. There are numerous technical difficulties with this method, but nevertheless, it is in principle possible to conduct such an experiment. Future research could examine if more effective protocols exist.
\\
\\
Our paper looked into the statistics of the signal that is created. Our results show that the statistics are indeed mixed, but do not follow thermal statistics. One noteworthy difference between the statistics of a thermal bath and the results obtained in our paper was the low frequency statistics. It can be shown that the $1/\omega_0^2$ dependence for low frequency leads to energy divergences. This is due to ultra-low frequency delays, which cannot be achieved in practice.
\\
\\
The issue regarding the infinite energy was also encountered for the case of uniformly accelerated mirror  \cite{Daiqin2016}. It is noted that the energy flux and particle flux of a uniformly accelerating mirror away from the horizon is actually zero \cite{PCWDavies1975, Birrell1982, Fulling1976, Grove1986}. The particles and energy are only created when there is a change in acceleration. For an eternally accelerated mirror, the radiation source can be traced back to the horizon, where there is a divergence in energy flux \cite{Frolov1979, Frolov1980, Frolov1999}. In our case, we make similar argument and argue that the divergence occurs due to accelerating the time delay source for an infinite time. 
\\
\\
Another intriguing motivation for studying the time-delay is a possible connection to the results presented in recent experiments by Riek et al \cite{Subcycle}. These authors measured the effect of a rapidly varying time-delay produced by transmission through a crystal with a changing refractive index. Due to the similarity between a time varying refractive index and acceleration \cite{refractiveindex}, we believe that analysing the effect of the accelerated time-delay may give further insight into the results obtained in this paper and lead to new experimental proposals. In the next section, we note some possible implications of our results on black-hole information paradox.
\subsection{A comment on black hole information paradox}
 The black hole information information paradox, \cite{Hawking1976, Susskind1993,Stephens1994, Mathur2005, Hawking2016, VBaccetti2016,Braunstein2013, Almheiri2013} points out the apparent contradiction between quantum mechanics and Hawking radiation. To restore the purity of the final state of an evaporated black hole, there must be hidden correlations in the final state. Some of the previous proposal were correlations between early and late time thermal bath \cite{Susskind1993,Stephens1994,Almheiri2013} and correlations between the thermal bath and curvature of space-time \cite{Nomura2016, Nomura2015a, Nomura2015b}. Our study raises the possibility that the correlations may exist between the distorted vacuum fluctuations. The equivalence principle ties a strong connection between gravity and acceleration \cite{Gravitation1973, Davies1977, Walker1985, Carlitz1987, Hawking1975}. Thus, we conjecture that the notion of apparent decoherence in the Rindler case can also be applied to the case of a black hole.

\clearpage
\begin{widetext}
\section*{Appendix}
\vspace{0.5cm}
\section{Correlation between Unruh operators due to Accelerated time delay/evolution}
The vacuum expectation values of the product of the two output Unruh operator are calculated by utilizing equation (35) and the fact that the Unruh and Minkowski vacuum coincides:
\begin{equation}
\begin{aligned}
\braket{0_M|\hat{c}_{\omega}^{\dag}{}'\hat{c}_{\omega'}{}'|0_M}&= F_1(\omega) \delta(\omega-\omega')
\\
\braket{0_M|\hat{d}_{\omega}^{\dag}{}'\hat{d}_{\omega'}{}'|0_M}&=F_1(\omega) \delta(\omega-\omega')
\\
\braket{0_M|\hat{c}_{\omega}{}'\hat{c}_{\omega'}{}'|0_M}&=0\\
\braket{0_M|\hat{d}_{\omega}{}'\hat{d}_{\omega'}{}'|0_M}&=0
\\
\braket{0_M|\hat{c}_{\omega}{}'\hat{d}_{\omega'}{}'|0_M}&=F_2(\omega) \delta(\omega-\omega')
\\
\braket{0_M|\hat{c}_{\omega}^{\dag}{}'\hat{d}_{\omega'}{}'|0_M}&=0
\end{aligned}
\end{equation}
where we have defined the following:
\begin{equation}
\begin{aligned}
F_1(\omega) & \equiv \cosh(r_{\omega})^2 \sinh(r_{\omega})^2(2-2 \cos(\omega \Delta))
\\
F_2(\omega)& \equiv \cosh(r_{\omega}) \sinh(r_{\omega}) (1- e^{i \omega \Delta})(\cosh(r_{\omega})^2 e^{-i \omega \Delta}-\sinh(r_{\omega})^2)
\end{aligned}
\end{equation}
All other combination can be found utilizing commutation relations for the Unruh operators or applying the complex conjugate.  
\end{widetext}
\section{Practical and Ideal Self Homodyne Detection}\label{sec: Reference}
\subsection{Practical Measurements}
We note that, if we explicitly calculate the quadrature amplitude, equation (36) without ignoring the terms which are in the order of $1/\sqrt{\alpha}$, this is the expression we obtain:
\begin{equation}
\begin{aligned}
X(\phi)&=\frac{1}{|\alpha|} \frac{\int \mathrm{d}\omega \; \delta(0) F_1(\omega)  }{\sqrt{1+2I_s}}
\end{aligned}
\end{equation}
In this case, the approximation that the above expression is 0 is not valid. As there are infinite particles, we require $|\alpha|$ to be infinite, which cannot be achieved in an experiment. This practicality issue will be addressed in this section. The results obtained in the paper are the idealized results which neglected these values. This issue also appears in the calculations for the variance as well.
\\
\\
In practice, we cannot measure infinitely small and large wavelength particles. This is due to the limitations caused by our experimental apparatus and set up. We assume that our detector can only measure frequencies between $k_{min}$ and $k_{max}$. We introduce a new subscript $Pr$ to denote practical measurements with low and high frequency cut-off. This can be compared with the subscript $Id$ which denotes the ideal measurements that were obtained in the paper.
\\
\\
The particle count measured by a practical detector is modelled by the operator defined in equation (61). 
\begin{widetext}
\begin{equation}
\begin{aligned}
\hat{N}_{Pr}& \equiv \int_{k_{min}}^{k_{max}} \mathrm{d}k \; \hat{e}_k^{\dag}{}''\hat{e}_k''
\\ &=\int \mathrm{d}\omega \mathrm{d}\omega' \; A_{\omega\omega',1}[\hat{c}_{\omega}^{\dag}{}''\hat{c}_{\omega'}{}''+\hat{d}_{\omega'}^{\dag}{}''\hat{d}_{\omega}{}'']+[A_{\omega\omega',2}\hat{c}_{\omega}^{\dag}{}''\hat{d}_{\omega'}{}''+A_{\omega\omega',2}^*\hat{d}_{\omega}^{\dag}{}''\hat{c}_{\omega'}{}'']
\\A_{\omega\omega',1} & \equiv \int\mathrm{d}k A_{k\omega}^*A_{k\omega'}, \;
A_{\omega\omega',2} \equiv \int\mathrm{d}k A_{k\omega}^*A_{k\omega'}^*
\end{aligned}
\end{equation}
The corresponding expectation value are calculated as follows:
\begin{equation}
\begin{aligned}
\braket{\hat{N}_{Pr}}  = & \int \mathrm{d}\omega \mathrm{d}\omega' \; A_{\omega\omega',1}[\braket{\hat{c}_{\omega}^{\dag}{}''\hat{c}_{\omega'}{}''}+\braket{\hat{d}_{\omega'}^{\dag}{}''\hat{d}_{\omega}{}''}]+2 Re[A_{\omega\omega',2}\braket{\hat{c}_{\omega}^{\dag}{}''\hat{d}_{\omega'}{}''}]
\\ = & \;|\alpha|^2 \int \mathrm{d}\omega \mathrm{d}\omega' \;  A_{\omega\omega',1}[(\cosh(r_{\omega})\cosh(r_{\omega'})+\sinh(r_{\omega})\sinh(r_{\omega'}) )g(\omega)g(\omega')^* )]
\\& \; \;\;\;\;\;\;\;\;\;\;\;\;\;\;\;\;\;\;\; - 2 Re[ A_{\omega\omega',2} \cosh(r_{\omega}) \sinh(r_{\omega})g(\omega) g(\omega')\alpha^2]
\\ & +  |\alpha|^0 \int \mathrm{d}\omega \; F_1(\omega) A_{\omega\omega,1}
\\
\braket{\hat{N}_{Pr,0}} = & \;|\alpha|^2 \int \mathrm{d}\omega \mathrm{d}\omega' \;  A_{\omega\omega',1}[(\cosh(r_{\omega})\cosh(r_{\omega'})+\sinh(r_{\omega})\sinh(r_{\omega'}) )g(\omega)g(\omega')^* )]
\\& \; \;\;\;\;\;\;\;\;\;\;\;\;\;\;\;\;\;\;\; - 2 Re[ A_{\omega\omega',2} \cosh(r_{\omega}) \sinh(r_{\omega'})g(\omega) g(\omega')\alpha^2]
\end{aligned}
\end{equation}
\end{widetext}
Thus, the practical quadrature amplitude is:
\begin{equation}
\begin{aligned}
X_{Pr}(\phi) &= \frac{\frac{1}{2\pi}\log (\frac{k_{max}}{k_{min}})\int \mathrm{d}\omega \; F_1(\omega)}{\sqrt{\braket{\hat{N}_0}}}
\\ &\approx 0
\end{aligned}
\end{equation}
The approximation is valid as $\int \mathrm{d}\omega \; F_1(\omega)$ is finite. How large $|\alpha|$ must be for the approximation to be valid will be analysed later in the appendix. We are now interested in calculating the variance. To do this, we must calculate the expectation value of $\hat{N}_{Pr}^2$. $\hat{N}_{Pr}^2$ is defined as follows:
\begin{widetext}
\begin{equation}
\begin{aligned}
\hat{N}_{Pr}{}^2 =  \int \mathrm{d}\omega \mathrm{d}\omega'\mathrm{d}\omega'' \mathrm{d}\omega''' &\; (A_{\omega\omega',1}A_{\omega''\omega''',1}[\hat{c}_{\omega}^{\dag}{}'\hat{c}_{\omega'}{}'\hat{c}_{\omega''}^{\dag}{}'\hat{c}_{\omega'''}{}'+\hat{d}_{\omega'}^{\dag}{}'\hat{d}_{\omega}{}'\hat{d}_{\omega'''}^{\dag}{}'\hat{d}_{\omega''}{}']+\{\hat{d}_{\omega'}^{\dag{}'}\hat{d}_{\omega}{}'\hat{c}_{\omega''}^{\dag}{}'\hat{c}_{\omega'''}{}'+\hat{c}_{\omega}^{\dag}{}'\hat{c}_{\omega'}{}'\hat{d}_{\omega'''}^{\dag}{}'\hat{d}_{\omega''}{}'\}]
\\
&+ A_{\omega\omega',2}A_{\omega''\omega''',2}^*\hat{c}_{\omega}^{\dag}{}'\hat{d}_{\omega'}{}'\hat{d}_{\omega''}^{\dag}{}'\hat{c}_{\omega'''{}'}+A_{\omega\omega',2}^*A_{\omega''\omega''',2}\hat{d}_{\omega}^{\dag}{}'\hat{c}_{\omega'}{}'\hat{c}_{\omega''}^{\dag}{}'\hat{d}_{\omega'''}{}'
\\
&+\{A_{\omega\omega',2}A_{\omega''\omega''',2}\hat{c}_{\omega}^{\dag}{}'\hat{d}_{\omega'}{}'\hat{c}_{\omega''}^{\dag}{}'\hat{d}_{\omega'''}{}'+A_{\omega\omega',2}^*A_{\omega''\omega''',2}^*\hat{d}_{\omega}^{\dag}{}'\hat{c}_{\omega'}{}'\hat{d}_{\omega''}^{\dag}{}'\hat{c}_{\omega'''}{}'\}
\\
&+\{A_{\omega\omega',1}A_{\omega''\omega''',2}\hat{c}_{\omega}^{\dag}{}'\hat{c}_{\omega'}{}'\hat{c}_{\omega''}^{\dag}{}'\hat{d}_{\omega'''}{}'+A_{\omega\omega',2}^*A_{\omega''\omega''',1}\hat{d}_{\omega}^{\dag}{}'\hat{c}_{\omega'}{}'\hat{c}_{\omega''}^{\dag}{}'\hat{c}_{\omega'''}{}'\}
\\
&+\{A_{\omega\omega',2}A_{\omega''\omega''',1}\hat{c}_{\omega}^{\dag}{}'\hat{d}_{\omega'}{}'\hat{c}_{\omega''}^{\dag}{}'\hat{c}_{\omega'''}{}'+A_{\omega\omega',1}A_{\omega''\omega''',2}^*\hat{c}_{\omega}^{\dag}{}'\hat{c}_{\omega'}{}'\hat{d}_{\omega''}^{\dag}{}'\hat{c}_{\omega'''}{}'\}
\\
&+ \{A_{\omega\omega',1}A_{\omega''\omega''',2}^*\hat{d}_{\omega'}^{\dag}{}'\hat{d}_{\omega}{}'\hat{d}_{\omega''}^{\dag}{}'\hat{c}_{\omega'''}{}'+A_{\omega\omega',2}A_{\omega''\omega''',1}\hat{c}_{\omega}^{\dag}{}'\hat{d}_{\omega'}{}'\hat{d}_{\omega'''}^{\dag}{}'\hat{d}_{\omega''}{}'\}
\\
&+\{A_{\omega\omega',2}^*A_{\omega''\omega''',1}\hat{d}_{\omega}^{\dag}{}'\hat{c}_{\omega'}{}'\hat{d}_{\omega'''}^{\dag}{}'\hat{d}_{\omega''}{}'+A_{\omega\omega',1}A_{\omega''\omega''',2}\hat{d}_{\omega'}^{\dag}{}'\hat{d}_{\omega}{}'\hat{c}_{\omega''}^{\dag}{}'\hat{d}_{\omega'''}{}'\}
\end{aligned}
\end{equation}
In the equation above, we have grouped the operators which are related by relabelling and a complex conjugate. The expectation value of this operator can be calculated by first computing the following correlation functions:
\begin{equation}
\begin{aligned}
\braket{\hat{c}_{\omega}^{\dag}{}''\hat{c}_{\omega'}{}''\hat{c}_{\omega''}^{\dag}{}''\hat{c}_{\omega'''}{}''}=\braket{\hat{c}_{\omega}^{\dag}{}''\hat{c}_{\omega'}{}''}\braket{\hat{c}_{\omega''}^{\dag}{}''\hat{c}_{\omega'''}{}''}&+g_c(\omega')g_c(\omega'')^*\delta(\omega-\omega''')F_1(\omega)
\\&+ g_c(\omega)^*g_c(\omega''')\delta(\omega'-\omega'')(1+F_1(\omega'))
\\&+|\alpha|^0\delta(\omega-\omega''')\delta(\omega'-\omega'')F_1(\omega)(1+F_1(\omega'))
\\
\braket{\hat{d}_{\omega}^{\dag}{}''\hat{d}_{\omega'}{}''\hat{d}_{\omega''}^{\dag}{}''\hat{d}_{\omega'''}{}''}=\braket{\hat{d}_{\omega}^{\dag}{}''\hat{d}_{\omega'}{}''}\braket{\hat{d}_{\omega''}^{\dag}{}''\hat{d}_{\omega'''}{}''}
&+g_d(\omega')g_d(\omega'')^*\delta(\omega-\omega''')F_1(\omega)
\\&+g_d(\omega)^*g_d(\omega''')\delta(\omega'-\omega'')(1+F_1(\omega'))
\\&+|\alpha|^0\delta(\omega-\omega''')\delta(\omega'-\omega'')F_1(\omega)(1+F_1(\omega'))
\\
\braket{\hat{c}_{\omega}^{\dag}{}''\hat{c}_{\omega'}{}''\hat{d}_{\omega''}^{\dag}{}''\hat{d}_{\omega'''}{}''}=\braket{\hat{c}_{\omega}^{\dag}{}''\hat{c}_{\omega'}{}''}\braket{\hat{d}_{\omega''}^{\dag}{}''\hat{d}_{\omega'''}{}''}
&+g_c(\omega)^*g_d(\omega'')^*\delta(\omega'-\omega''')F_2(\omega')
\\&+g_c(\omega')g_d(\omega''')\delta(\omega-\omega'')F_2(\omega)^*
\\&+|\alpha|^0 \delta(\omega-\omega'')\delta(\omega'-\omega''')F_2(\omega)^* F_2(\omega')
\\
\end{aligned}
\end{equation}
\begin{equation}
\begin{aligned}
\braket{\hat{c}_{\omega}^{\dag}{}''\hat{d}_{\omega'}{}''\hat{d}_{\omega''}^{\dag}{}''\hat{c}_{\omega'''}{}''}=\braket{\hat{c}_{\omega}^{\dag}{}''\hat{d}_{\omega'}{}''}\braket{\hat{d}_{\omega''}^{\dag}{}''\hat{c}_{\omega'''}{}''}
&+g_c(\omega)^*g_c(\omega''')\delta(\omega'-\omega'')(1+F_1(\omega'))
\\&+g_d(\omega')g_d(\omega'')^*\delta(\omega-\omega''')F_1(\omega)
\\&+g_c(\omega)^*g_d(\omega'')^*\delta(\omega'-\omega''')F_2(\omega')
\\&+g_d(\omega')g_c(\omega''')\delta(\omega-\omega'')F_2(\omega)^*
\\&+|\alpha|^0 \delta(\omega-\omega'')\delta(\omega'-\omega''')F_2(\omega') F_2(\omega)^*
\\&+|\alpha|^0\delta(\omega-\omega''')\delta(\omega'-\omega'')F_1(\omega)(1+F_1(\omega'))
\\
\braket{\hat{d}_{\omega}^{\dag}{}''\hat{c}_{\omega'}{}''\hat{c}_{\omega''}^{\dag}{}''\hat{d}_{\omega'''}{}''}=\braket{\hat{d}_{\omega}^{\dag}{}''\hat{c}_{\omega'}{}''}\braket{\hat{c}_{\omega''}^{\dag}{}''\hat{d}_{\omega'''}{}''}
&+g_c(\omega')g_c(\omega'')^* \delta(\omega-\omega''')F_1(\omega)
\\&+g_d(\omega)^*g_d(\omega''')\delta(\omega'-\omega'')(1+F_1(\omega'))
\\&+g_d(\omega)^*g_c(\omega'')^*\delta(\omega'-\omega''')F_2(\omega')
\\&+g_c(\omega')g_d(\omega''')\delta(\omega-\omega'')F_2(\omega)^*
\\&+|\alpha|^0 \delta(\omega-\omega'')\delta(\omega'-\omega''')F_2(\omega')F_2(\omega)^*
\\&+|\alpha|^0 \delta(\omega'-\omega'')\delta(\omega-\omega''')F_1(\omega)(1+F_1(\omega'))
\\
\braket{\hat{c}_{\omega}^{\dag}{}''\hat{d}_{\omega'}{}''\hat{c}_{\omega''}^{\dag}{}''\hat{d}_{\omega'''}{}''}=\braket{\hat{c}_{\omega}^{\dag}{}''\hat{d}_{\omega'}{}''}\braket{\hat{c}_{\omega''}^{\dag}{}''\hat{d}_{\omega'''}{}''}&
\\
\braket{\hat{c}_{\omega}^{\dag}{}''\hat{c}_{\omega'}{}''\hat{c}_{\omega''}^{\dag}{}''\hat{d}_{\omega'''}{}''}=\braket{\hat{c}_{\omega}^{\dag}{}''\hat{c}_{\omega'}{}''}\braket{\hat{c}_{\omega''}^{\dag}{}''\hat{d}_{\omega'''}{}''}
&+g_c(\omega)^*g_c(\omega'')^* \delta(\omega'-\omega''') F_2(\omega')
\\
&+g_c(\omega)^*g_d(\omega''')\delta(\omega'-\omega'')(F_1(\omega')+1)
\\
\braket{\hat{c}_{\omega}^{\dag}{}''\hat{d}_{\omega'}{}''\hat{c}_{\omega''}^{\dag}{}''\hat{c}_{\omega'''}{}''}=\braket{\hat{c}_{\omega}^{\dag}{}''\hat{d}_{\omega'}{}''}\braket{\hat{c}_{\omega''}^{\dag}{}''\hat{c}_{\omega'''}{}''}&+g_c(\omega)^*g_c(\omega'')^* \delta(\omega'-\omega'') F_2(\omega')
\\
&+g_d(\omega')g_c(\omega'')^*\delta(\omega-\omega''')F_1(\omega)
\\
\braket{\hat{d}_{\omega}^{\dag}{}''\hat{d}_{\omega'}{}''\hat{d}_{\omega''}^{\dag}{}''\hat{c}_{\omega'''}{}''}
=\braket{\hat{d}_{\omega}^{\dag}{}''\hat{d}_{\omega'}{}''}\braket{\hat{d}_{\omega''}^{\dag}{}''\hat{c}_{\omega'''}{}''}
&+g_d(\omega)^*g_d(\omega'')^* \delta(\omega'-\omega''') F_2(\omega')
\\
&+g_d(\omega)^*g_c(\omega''')\delta(\omega'-\omega'')(F_1(\omega')+1)
\\
\braket{\hat{d}_{\omega}^{\dag}{}''\hat{c}_{\omega'}{}''\hat{d}_{\omega''}^{\dag}{}''\hat{d}_{\omega'''}{}''}=\braket{\hat{d}_{\omega}^{\dag}{}''\hat{c}_{\omega'}{}''}\braket{\hat{d}_{\omega''}^{\dag}{}''\hat{d}_{\omega'''}{}''}
&+g_d(\omega)^*g_d(\omega'')^* \delta(\omega'-\omega''') F_2(\omega')
\\
&+g_c(\omega')g_d(\omega'')^*\delta(\omega-\omega''')F_1(\omega))
\end{aligned}
\end{equation}
All other expressions can be found by applying a complex conjugate to the expressions above or by utilizing the fact that $\hat{c}_{\omega}$ commutes with $\hat{d}_{\omega}$. We introduce the G functions to simplify further calculations:
\begin{equation}
\begin{aligned}
G_{\alpha \beta \gamma \delta}(\omega,\omega',\omega'',\omega''') &\equiv \braket{\hat{\alpha}_{\omega}^{\dag}{}''\hat{\beta}_{\omega'}{}''\hat{\gamma}_{\omega''}^{\dag}{}''\hat{\delta}_{\omega'''}{}''}-\braket{\hat{\alpha}_{\omega}^{\dag}{}''\hat{\beta}_{\omega'}{}''}\braket{\hat{\gamma}_{\omega''}^{\dag}{}''\hat{\delta}_{\omega'''{}''}}
\end{aligned}
\end{equation}
Where $\alpha,\beta,\gamma,\delta \in {c,d}$. The explicit expressions of these terms can be found by plugging in the expression written in equation (65) and (66).  We introduce a subscript to these G-functions: $G_{\alpha\beta\gamma\delta,n}$, where $n \in {0,2}$. The new subscript denotes the 0th order $\alpha$ term or the 2nd order $\alpha$ term. Utilizing equations (64) to (67), we find that the particle number fluctuation can be written in the following way:
\begin{equation}
\begin{aligned}
(\Delta\braket{\hat{N}_{Pr}})^2 =  \int \mathrm{d}\omega \mathrm{d}\omega'\mathrm{d}\omega'' \mathrm{d}\omega''' &\; A_{\omega\omega',1}A_{\omega''\omega''',1}G_{cccc}(\omega,\omega',\omega'',\omega''')+A_{\omega\omega',1}^*A_{\omega''\omega''',1}^*G_{dddd}(\omega,\omega',\omega'',\omega''')
\\ & + 2A_{\omega,\omega',1}A_{\omega'',\omega''',1}^*G_{ccdd}(\omega,\omega',\omega'',\omega''')
\\ &+A_{\omega\omega',2}A_{\omega''\omega''',2}^*G_{cddc}(\omega,\omega',\omega'',\omega''') +A_{\omega\omega',2}^*A_{\omega''\omega''',2}G_{dccd}(\omega,\omega',\omega'',\omega''')
\\
&+2Re [A_{\omega\omega',1}^*A_{\omega''\omega''',2}^*G_{cccd}(\omega,\omega',\omega'',\omega''')^*+A_{\omega\omega',2}^* A_{\omega''\omega''',1}^* G_{cdcc}(\omega,\omega',\omega'',\omega''')^*
\\
&\;\;\;\;\;\;\;\;\;\;+ A_{\omega\omega',1}A_{\omega''\omega''',2}^* G_{dddc}(\omega,\omega',\omega'',\omega''')+A_{\omega\omega',2}^*A_{\omega''\omega''',1}G_{dcdd}(\omega,\omega',\omega'',\omega''')]
\end{aligned}
\end{equation}
We notice that every term in the last two lines are proportional to $\alpha^2$. Following similar steps to the paper, we write the variance of the signal in a compact way:
\begin{equation}
\begin{aligned}
V_{Pr}(\phi) & = \frac{(\Delta\braket{\hat{N_{Pr}}})^2}{\braket{\hat{N}_{Pr,0}}}
\\ &=V_{1,2}' + V_{2}' \cos(\theta-2\phi)+ V_{1,0}'
\\ & \approx V_{1,2}' + V_{2}' \cos(\theta-2\phi)
\end{aligned}
\end{equation}
Where we have defined the following:
\begin{equation}
\begin{aligned}
V_{1,n}'\equiv \frac{1}{\braket{\hat{N}_0}}\int \mathrm{d}\omega \mathrm{d}\omega'\mathrm{d}\omega'' \mathrm{d}\omega''' &\; A_{\omega\omega',1}A_{\omega''\omega''',1}G_{cccc,n}(\omega,\omega',\omega'',\omega''')+A_{\omega\omega',1}^*A_{\omega''\omega''',1}^*G_{dddd,n}(\omega,\omega',\omega'',\omega''')
\\ & + 2A_{\omega,\omega',1}A_{\omega'',\omega''',1}^*G_{ccdd,n}(\omega,\omega',\omega'',\omega''')
\\ &+A_{\omega\omega',2}A_{\omega''\omega''',2}^*G_{cddc,n}(\omega,\omega',\omega'',\omega''') +A_{\omega\omega',2}^*A_{\omega''\omega''',2}G_{dccd,n}(\omega,\omega',\omega'',\omega''')
\\
\end{aligned}
\end{equation}
\begin{equation}
\begin{aligned}
V_{2,\phi}'\equiv \frac{2}{\braket{\hat{N}_0}}\int \mathrm{d}\omega \mathrm{d}\omega'\mathrm{d}\omega'' \mathrm{d}\omega''' &A_{\omega\omega',1}^*A_{\omega''\omega''',2}^*G_{cccd}(\omega,\omega',\omega'',\omega''')^*+A_{\omega\omega',2}^* A_{\omega''\omega''',1}^* G_{cdcc}(\omega,\omega',\omega'',\omega''')^*
\\
&\;\;\;\;\;\;\;\;\;\;+ A_{\omega\omega',1}A_{\omega''\omega''',2}^* G_{dddc}(\omega,\omega',\omega'',\omega''')+A_{\omega\omega',2}^*A_{\omega''\omega''',1}G_{dcdd}(\omega,\omega',\omega'',\omega''')
\end{aligned}
\end{equation}
\begin{equation}
\begin{aligned}
\overline{V}_2'&=|V_{2,\phi}'|
\\
e^{i \theta} & \equiv \frac{V_{2,\phi=0}'}{|V_{2,\phi=0}'|}
\end{aligned}
\end{equation}
\end{widetext} 
$V_{1,2}'$ can be interpreted as the average noise of the signal. $V_{2}'$ can be interpreted as the amount of squeezing in the signal. $V_{1,0}'$ can be interpreted as the error that arises due to the construction of self-homodyne detection. 
In the next section we will conduct numerical analysis of the quadrature amplitude and variance.
\subsection{Numerical Analysis of Practical Measurements}
\subsubsection{Particle Number}
In this section, we look into how various parameters affect the particle count of a coherent Rindler signal with an amplitude of $|\alpha|=1$. This section will analyse the necessary conditions for $\braket{\hat{N}_{Pr}} \approx \braket{\hat{N}_{Id}}$. This is important, as if this is not satisfied it would mean significant amount of the signal was traced out.
\\
\\
\begin{figure}[h!]
\includegraphics[width=0.45\textwidth]{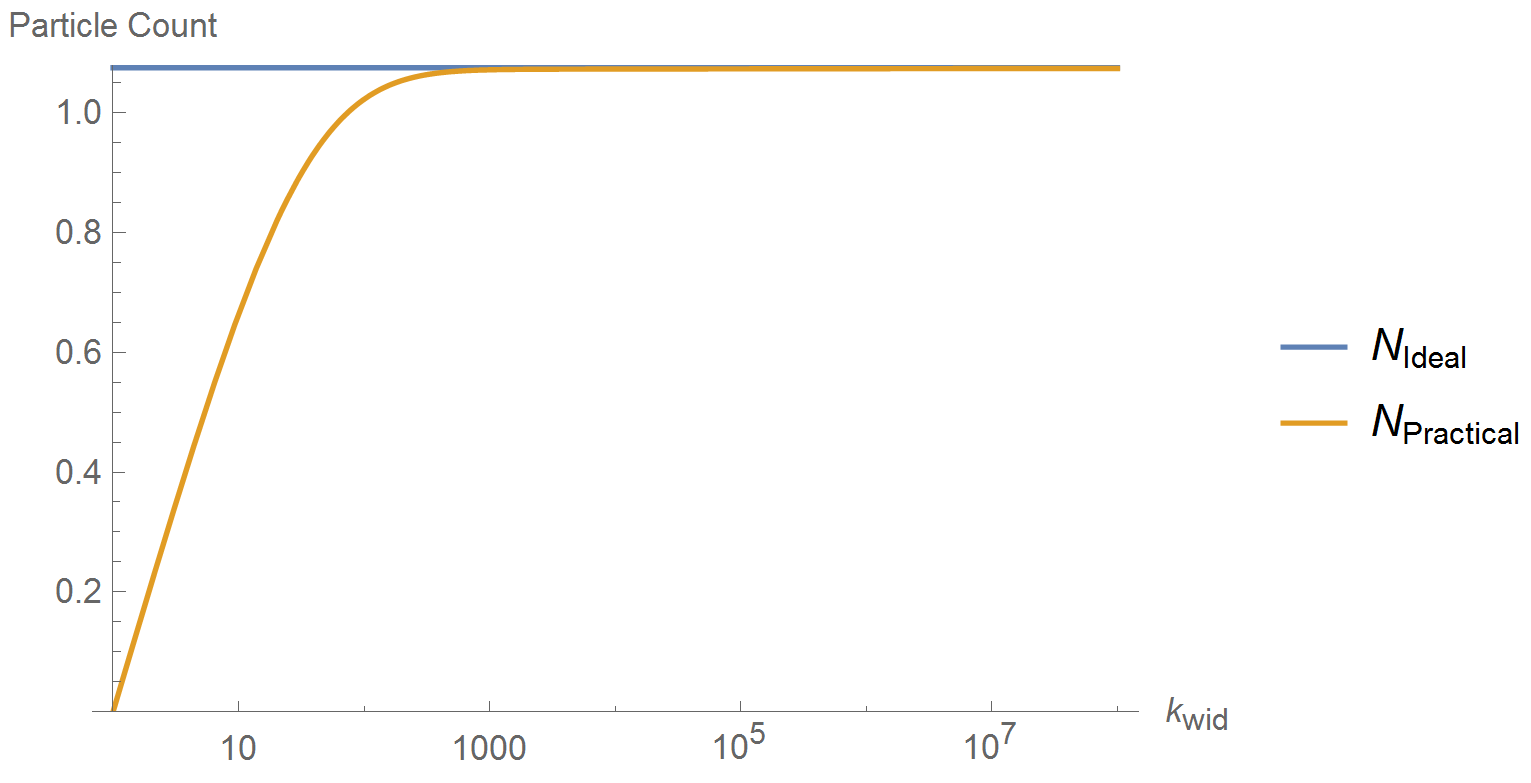}
\hfill
\caption{The graph above demonstrates how $k_{wid}$ affects the particle count detected by the Minkowski detector. We have utilized the following settings: $\omega_0=0.6, \; k_{med}=1, \; \delta = 0.4\omega_0, \; v_c=0.577, \; a=1$.}
\end{figure}
Fig. 6 demonstrates how $\braket{\hat{N}_{Pr}}$ converges towards $\braket{\hat{N}_{Id}}$ as we increase $k_{wid}$. We have defined the following: 
\begin{equation}
\begin{aligned}
k_{med} &\equiv \sqrt{k_{max}k_{min}}
\\
k_{wid} &\equiv \sqrt{k_{max}/k_{min}}
\end{aligned}
\end{equation}
\noindent We find that the particle count converges to the particle count $\braket{\hat{N}_{0,Id}}$ by increasing $k_{wid}$. 
To analyse the convergence rate, we now analyse how $\delta$ affects the particle count. The Rindler coordinate $v$ is related to the Minkowski coordinate $V$ in the following way:
\begin{equation}
\begin{aligned}
V=a^{-1}e^{-av} 
\end{aligned}
\end{equation}
From this equation, we conclude that a constant oscillation in the Rindler coordinate would result in a exponentially decaying frequency in the Minkowski coordinate. This ties a strong relationship bewteen the Rindler position and Minkowski frequency.
Utilizing this notion, and the fact that the field of the operator $\hat{a}_g$ is,
\begin{equation}
\begin{aligned}
f_{\hat{a}_g}(v)&= \left(\frac{1}{2\pi}\right)^{1/4} \sqrt{\frac{\delta}{\omega _0}} e^{-\delta^2 (v-v_c)^2}e^{-i\omega_0(v-v_c)}
\end{aligned}
\end{equation}
We conclude that the following condition should be satisfied to measure 2 standard deviation of the signal:
\begin{equation}
k_{wid}> a \;e^{\sqrt{2} a \delta^{-1}}
\end{equation}
Two standard deviation of a Gaussian covers $97.7 \%$ of the signal. As a result, we expect $\braket{\hat{N}_{Pr}}/\braket{\hat{N}_{Id}} \approx 0.977$ when $k_{wid} = a \;e^{\sqrt{2} a \delta^{-1}}$. We verify this conjecture through Fig. 7.
\\
\\
\begin{figure}[h!]
\subfigure[$\omega_0=0.5$]{\includegraphics[width=0.45\textwidth]{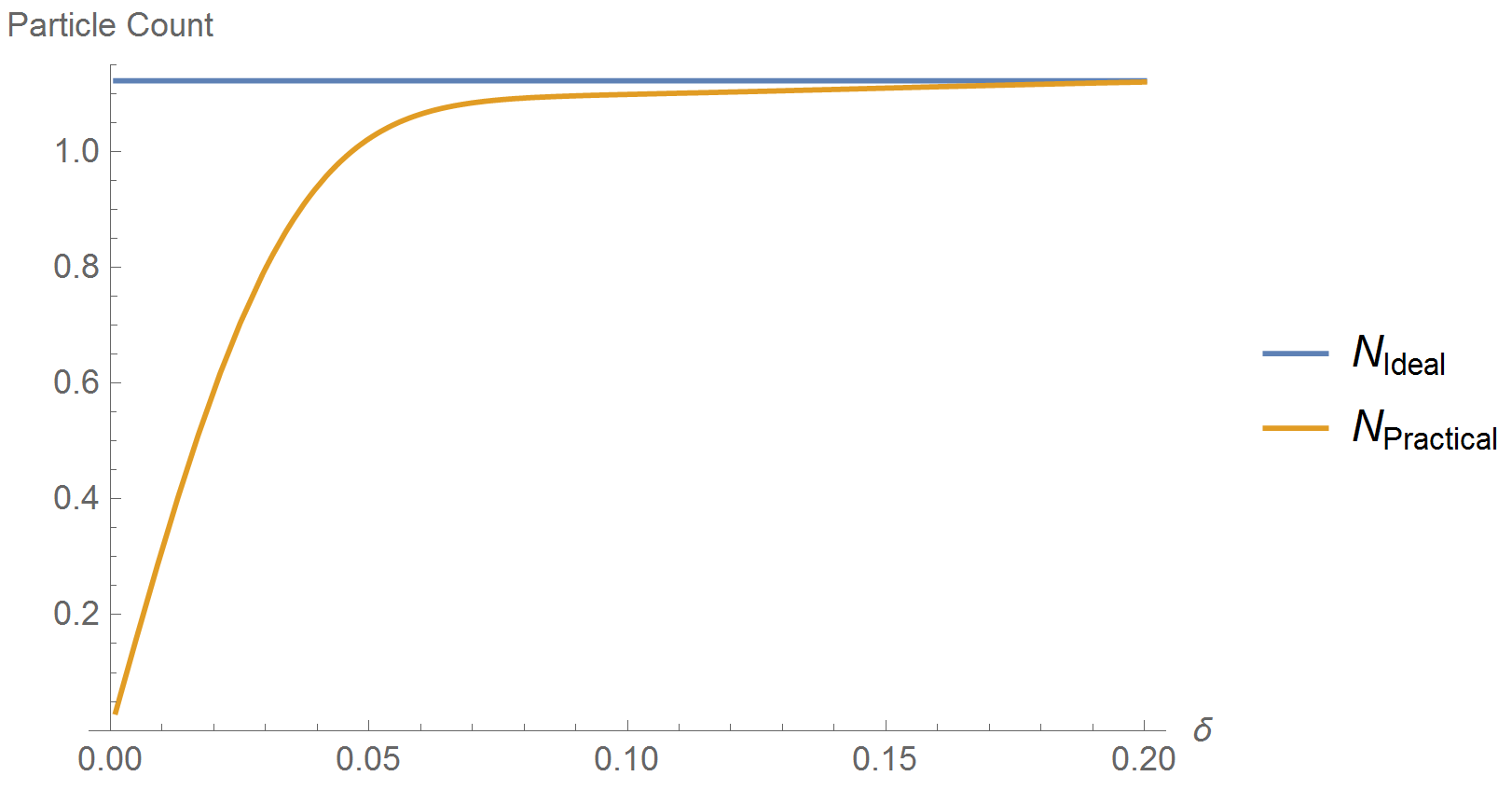}}
\hfill
\caption{The graph above demonstrates how $k_{width}$ affects the particle count detected by the Minkowski detector, for various $\omega_0$. We have utilized the following settings: $k_{med}=1, k_{wid}=10^8, v_c=0.577, a=1$.}
\end{figure}
From equation (76), it is found that when we have $k_{wid}=1 \times 10^8$, then 2 standard deviation of the signal is covered when $\delta \approx 0.077$. Further analysis shows that when $\delta=0.077$, $\braket{\hat{N}_{Pr}}/\braket{\hat{N}_{Id}} \approx 0.977$, regardless of $\omega_0$. This validates the conjecture made in equation (67). It is concluded that the spatial width of a Rindler signal has a very strong correlation with the frequency width of the signal in the Minkowski frame.
\\
\\
As we have tied a strong relation between the Rindler position and Minkowski frequency, there must be a connection between the central Rindler position $v_c$ and central Minkowski frequency $k_{med}$. We analyse the oscillatory behaviour within the integrand of $\braket{\hat{N}_{Pr}}$, found in equation (62). By utilizing the low frequency limit for the gamma function, $\Gamma(1+i x)\approx e^{-i \gamma x}$, we find that we can cancel out all of the oscillatory behaviour within the integrand by setting $v_c$ as follows:
\begin{equation}
v_c= \frac{1}{a}(0.577) + \log ([k_{med}]/a)
\end{equation}
Where $\gamma$ is the Euler constant. This expression explicitly demonstrates the connection between the Rindler position of the wave-packet mode, and the frequency in the Minkowski frame. 
\\
\\
How the particle count changes with $\omega_0$ is demonstrated in Fig. 8. It is found that the ideal and practical particle count coincides with each other for a smaller $\omega_0$ with larger $k_{wid}$.
\begin{figure}[h!]
\subfigure[$k_{wid}=10^6$]{\includegraphics[width=0.45\textwidth]{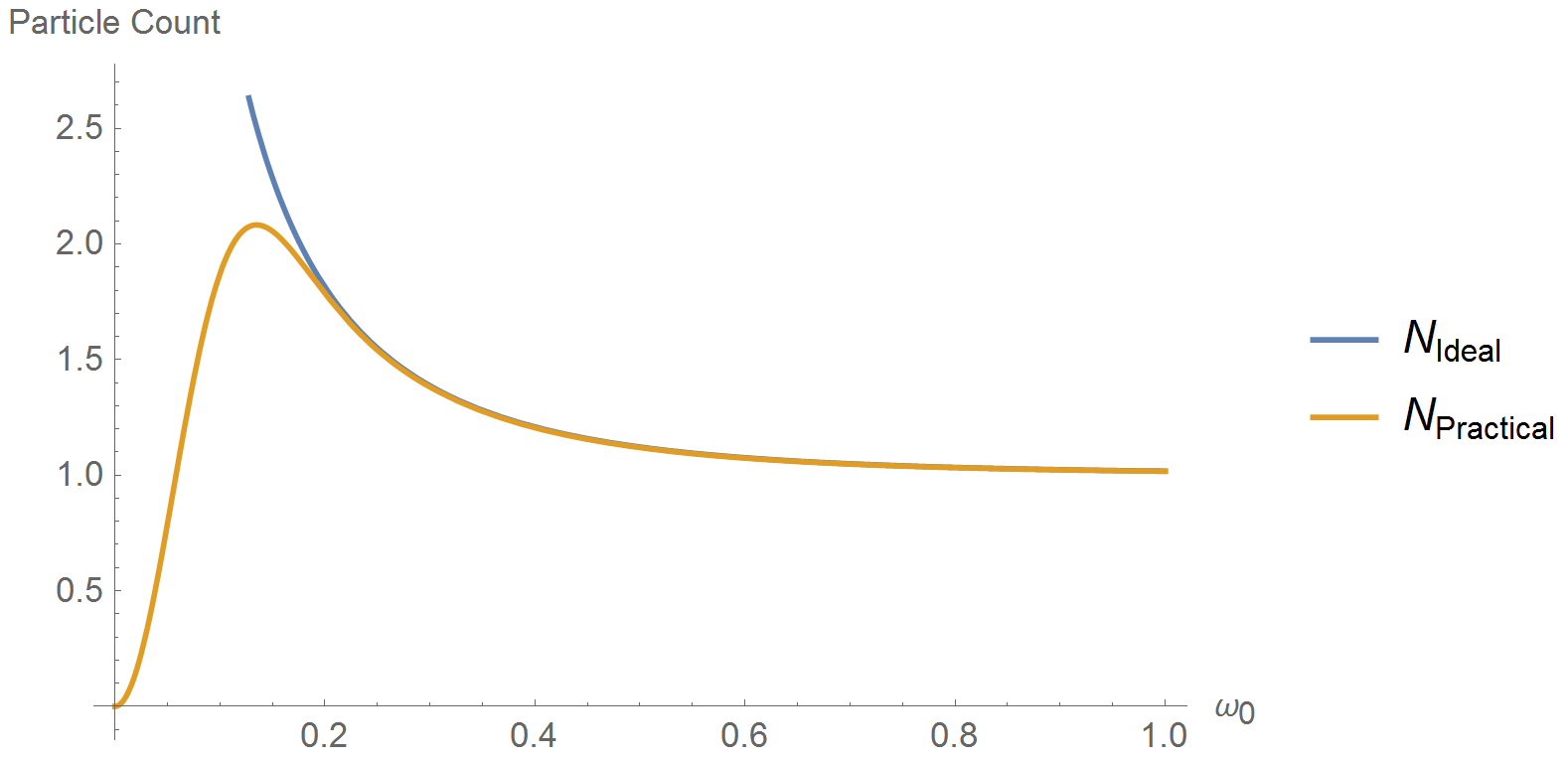}}
\hfill
\caption{The graphs above demonstrates how $\omega_0$ affects the particle count detected by the Minkowski detector. We have utilized the following settings: $k_{med}=1, \delta=0.4 \omega_0, v_c'=0.577, a=1$.}
\end{figure}
\subsubsection{Quadrature Amplitude}
In this section we look at $\braket{\hat{N}_{Pr}}-\braket{\hat{N}_{Pr,0}}$ and look at the validity of equation (63). The approximation made in this equation is valid when $\braket{\hat{N}_{Pr}}-\braket{\hat{N}_{Pr,0}} \ll (I_c+I_s)|\alpha|$. Thus, we analyse how large we must set $|\alpha|$ for the approximation in equation (63) to be valid. $\braket{\hat{N}_{Pr}}-\braket{\hat{N}_{Pr,0}}$ can be simplified as follows:
\begin{equation}
\begin{aligned}
\braket{\hat{N}_{Pr}} - \braket{\hat{N}_{Pr,0}} =\frac{1}{ \pi} \log(k_{wid}) \int \mathrm{d} \omega \; F_1(\omega ; \Delta)
\end{aligned}
\end{equation}
Looking at this equation, it is clear that the particle count is proportional to $\log(k_{wid})$. We analyse how $\Delta$ affects the particle count in Fig. 9.
\\
\\
\begin{figure} [h!]
\centering
\includegraphics[width=0.45\textwidth]{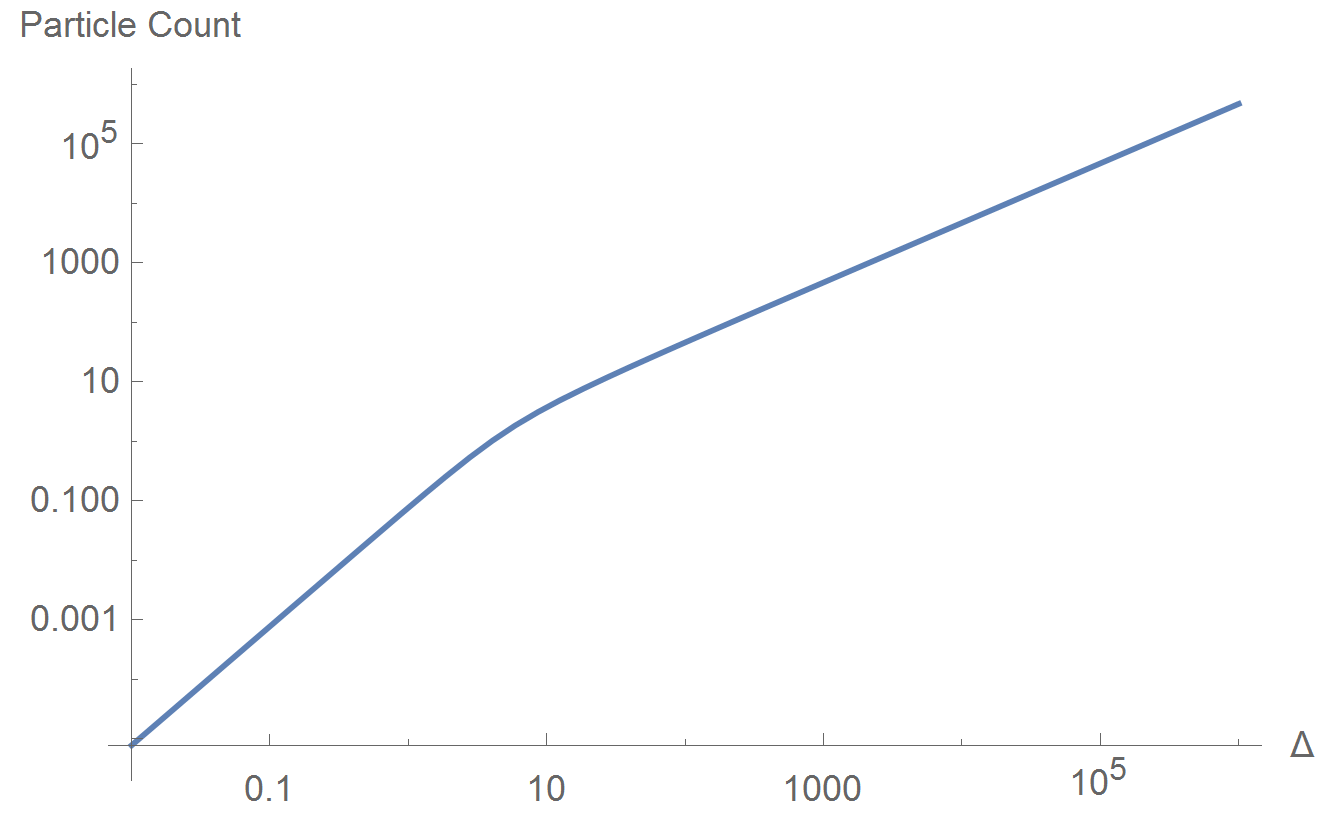}
\caption{The graph above is a plot of how the particle count, $\braket{\hat{N}_0} X(\phi)$, is affected by $\Delta$. We have utilized the following settings: $k_{wid}=10^8$}
\label{fig: Unitary}
\end{figure}
\noindent 
By analysing this graph, we find that the sufficient condition to assume $X(\phi) \approx 0$ is when the amplitude of the local oscillator satisfies the following:
\begin{equation}
|\alpha|\gg \Delta \frac{\log_{10} (k_{med})}{16}
\end{equation}
\subsubsection{Variance}
We now look at whether there are practical settings where the ideal and practical variances coincides with each other. This can be done by looking into the validity of the following equations:
\begin{align}
&V_{Pr}\approx V_{1,2}' + V_{2}'\cos(\theta-2\phi)
\\
&V_{1,2}' + V_{2}' \cos(\theta-2\phi)\approx V_{Id}
\end{align}
In this section we will explore the validity of the the latter equation. In the previous section we looked at the condition in which most of the coherent signal is observed. In this section we explore whether there are any further constraint for equation (81) to be valid.
\\
\\
We first examine how $\Delta$ affects the convergence bewteen practical and ideal variance. From Fig. 10 it is found that the practical variance coincides with the ideal case for $\omega_0=0.6$ regardless of $\Delta$. This is because we have set $k_{wid}=10^6$, which is large enough for more than 2 standard deviation of the signal to be measured. As a result, we conclude that $\Delta$ is not responsible for the relative deviation between the ideal and practical results. This makes sense, as $\Delta$ does not change which part of the signal is traced out. The effect of $\Delta$ on the variance will be discussed further in the following chapter. 
\\
\\
\begin{figure}[h!]
\hfill
\subfigure[$\omega_0=0.6$]{\includegraphics[width=0.45\textwidth]{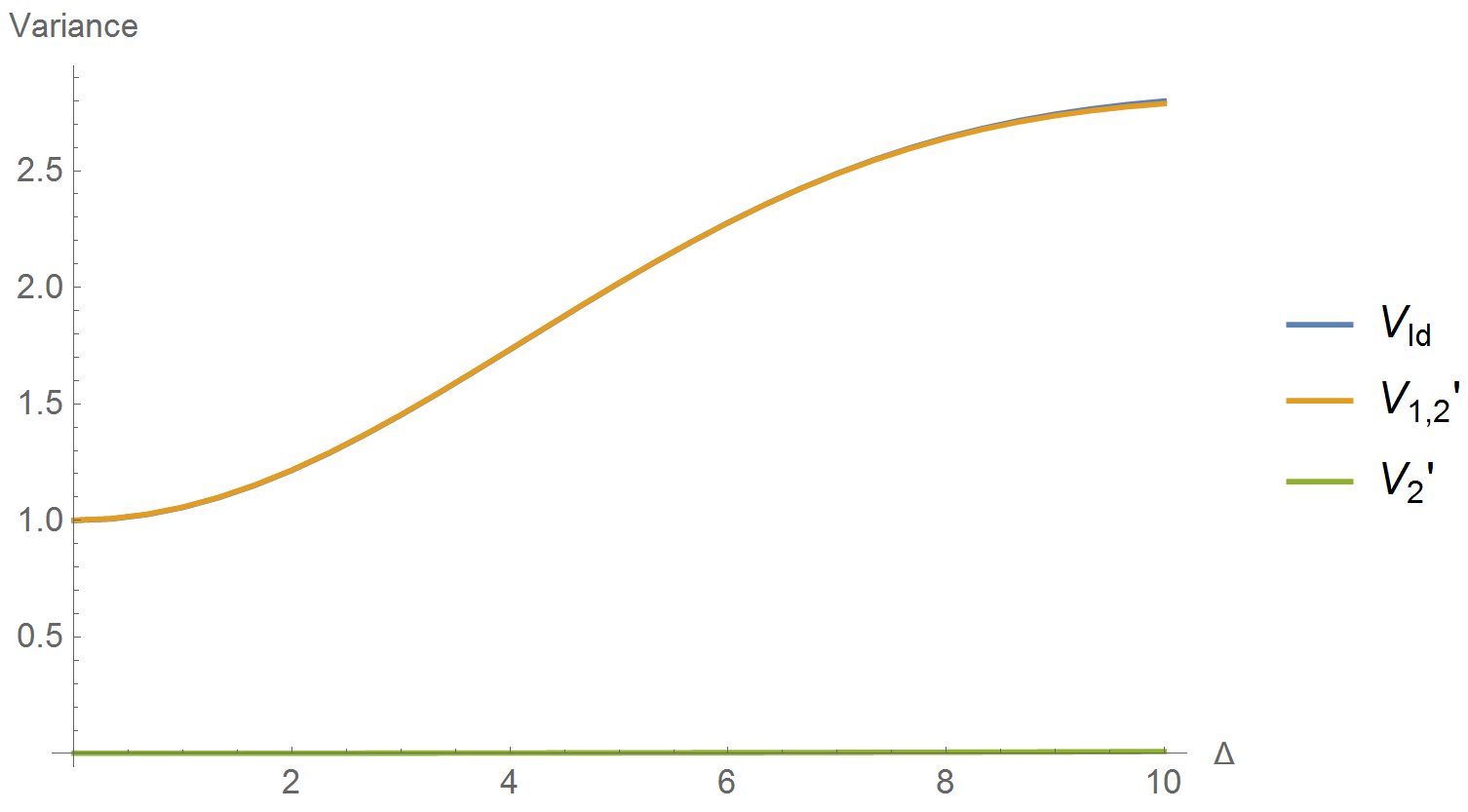}}
\caption{The graph above demonstrates how $\Delta$ affects the variance of the particle count detected by the Minkowski detector, for various $\omega_0$. We have utilized the following settings: $k_{med}=1, k_{wid}=10^6, \delta=0.4 \omega_0, v_c=0.577, a=1$.}
\end{figure}
\begin{figure} [h!]
\centering
\includegraphics[width=0.45\textwidth]{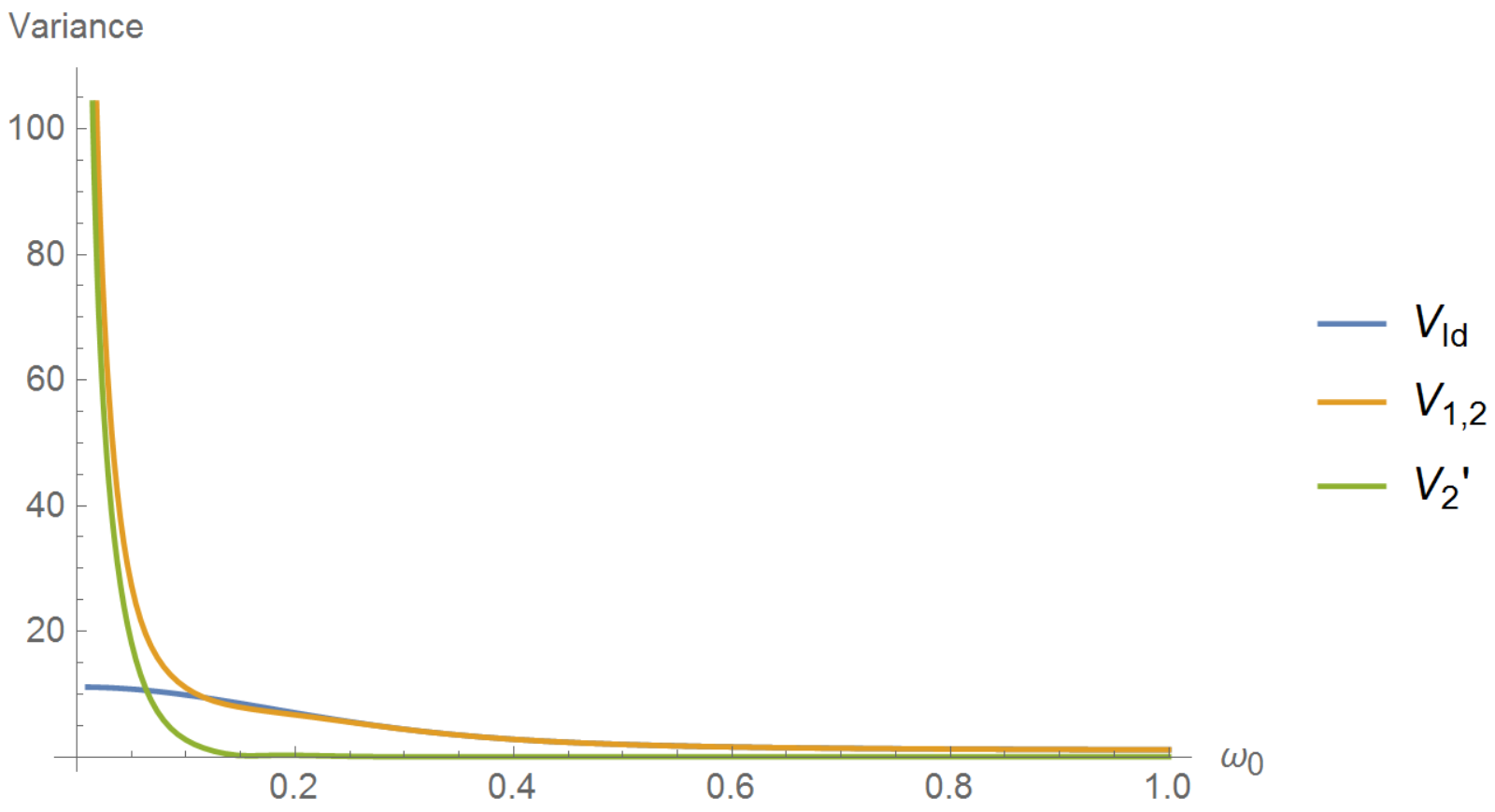}
\caption{The plots show how $\omega_0$ affects the variance. We have utilized the following settings: $k_{med}=1, k_{wid}=10^6, \delta=0.4 \omega_0, v_c'=0.577, a=1, \Delta=10$.}
\label{fig: Unitary}
\end{figure}
\\
\noindent We now examine how $\omega_0$ affects the variance. From Fig 11, we find that the variance follows a similar trend to what was observed for the particle count. The practical and ideal variance deviates from each other due to a $\delta$ that is too small compared to $k_{wid}$. It is interesting to note the squeezing effect that appears with smaller $\delta$. The squeezing effect appears when we introduce a low and high frequency cut-off. From this graph we can conclude that squeezing arises from tracing off important parts of the signal. Tracing off information not only causes mixing, but can also cause squeezing. Previously, it was shown that squeezing is observed from an accelerated mirror when the signal was analysed with reference to Minkowski frequencies \cite{Daiqin2016}. In this paper we showed that the squeezing effect observed in their paper is removed if we conduct self-homodyne detection with respect to Rindler frequencies. The squeezing observed in their paper was a result of tracing out correlations that existed between Unruh/Rindler modes.
\\
\\
In this section, we looked at the convergence rate of the variance and found that $\Delta$ does not play a huge role on the amount of error from the ideal case. We found that the error arises when important parts of the signal is traced out. The error is suppressed when equation (76) is satisfied. In the following section we will look into how large $|\alpha|$ must be in order to neglect the 0th order term.
\subsubsection{0th Order Variance Term}
In this section we examine the particle fluctuation, $\braket{\hat{N}_{Pr,0}} \times V_{1,0}$. By setting $|\alpha|^2 \gg \braket{\hat{N}_0}V_{1,0}$, equation (80) is satisfied. 
\\
\\
We first look into how $k_{wid}$ affects the particle fluctuations. Fig. 12 is a log-linear plot of particle fluctuation versus $k_{wid}$. This graph shows that the particle fluctuation is approximately logarithmically proportional to $k_{wid}$ for $k_{wid}$ larger than $10$.
\begin{figure} [h!]
\centering
\includegraphics[width=0.45\textwidth]{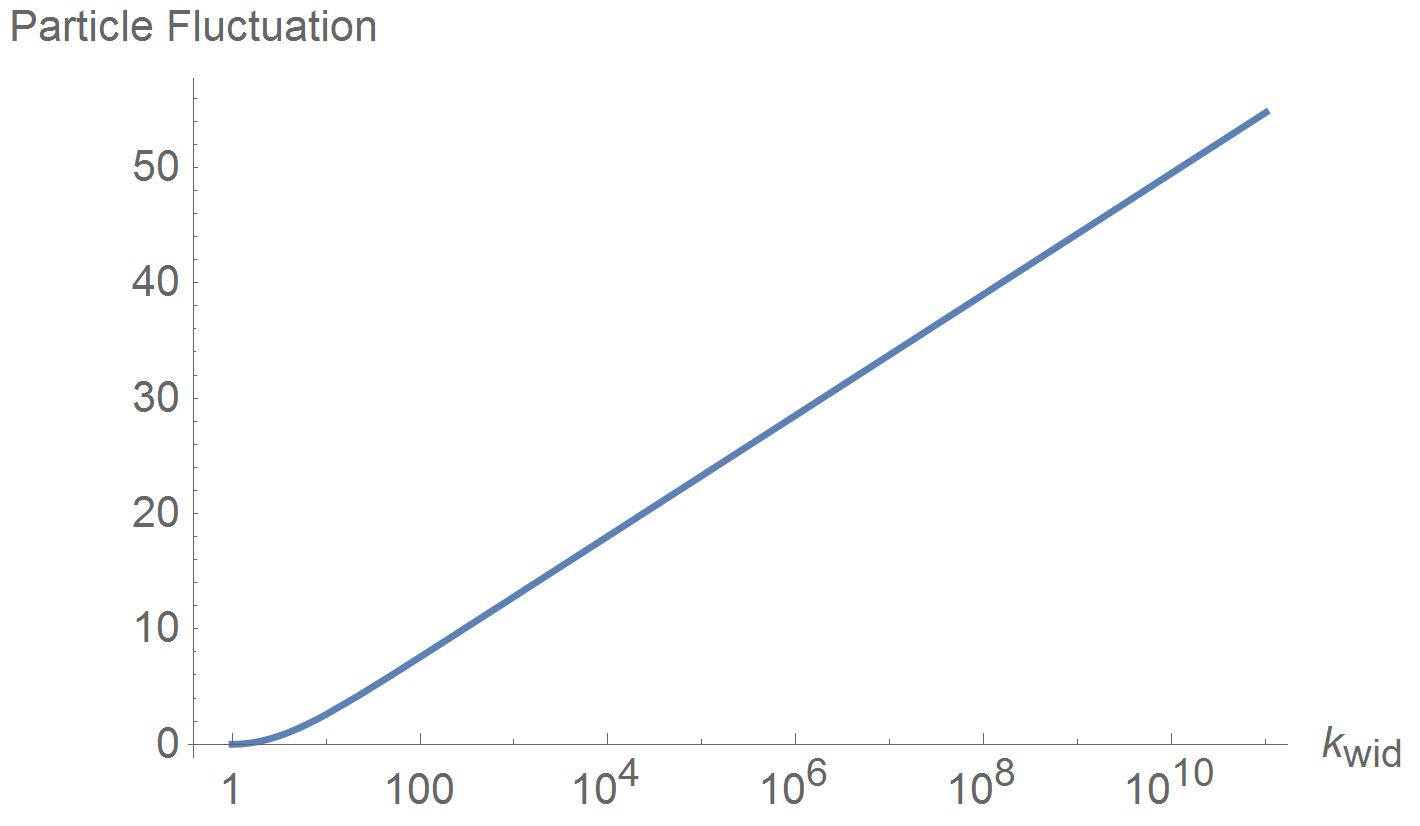}
\caption{The graph above is a plot of how the 0th order particle fluctuation, $\braket{\hat{N}_0}V_{1,0}'$, is affected by $k_{wid}$. We have utilized the following settings: $\Delta=10$}
\label{fig: Unitary}
\end{figure}
\\
\noindent We now look into how $\Delta$ affects the particle fluctuation. Fig. 13 is a log-log plot of the particle fluctuation versus $k_{wid}$. By a linear regression, we find that the particle fluctuation is quadratically proportional to the particle count. It is found that there is an increase in proportionality constant between $\Delta<10$ and $\Delta >100$. As we are interested in a sufficient condition to neglect the 0th order term, we consider the case when $\Delta > 100$. We find $\braket{\hat{N}_0}V_{1,0} \approx \Delta^2$. Combining this result with the result from Fig. 12, the sufficient condition to neglect the 0th order term is as follows:
 \begin{equation}
|\alpha|\gg \Delta \sqrt{\log(k_{wid})/6}
 \end{equation}
It is found that this condition puts a larger lower bound on $|\alpha|$ than equation (79) for $k_{wid}< 4 \times 10^{42}$. As it is impossible to reach this bound in a practical experiment, we conclude equation (82) must be satisfied for our experiment to neglect the 0th order term.
\\
\\
We have now demonstrated that there is a regime in which the practical measurement converges with the ideal measurement.  We showed that when equation (76) and (82) are satisfied, and with the correct $k_{mid}$, the practical measurement and the ideal measurements coincides with each other. This demonstrates the validity of the results found in equation (36) and (37).
\begin{figure} [h!]
\centering
\includegraphics[width=0.45\textwidth]{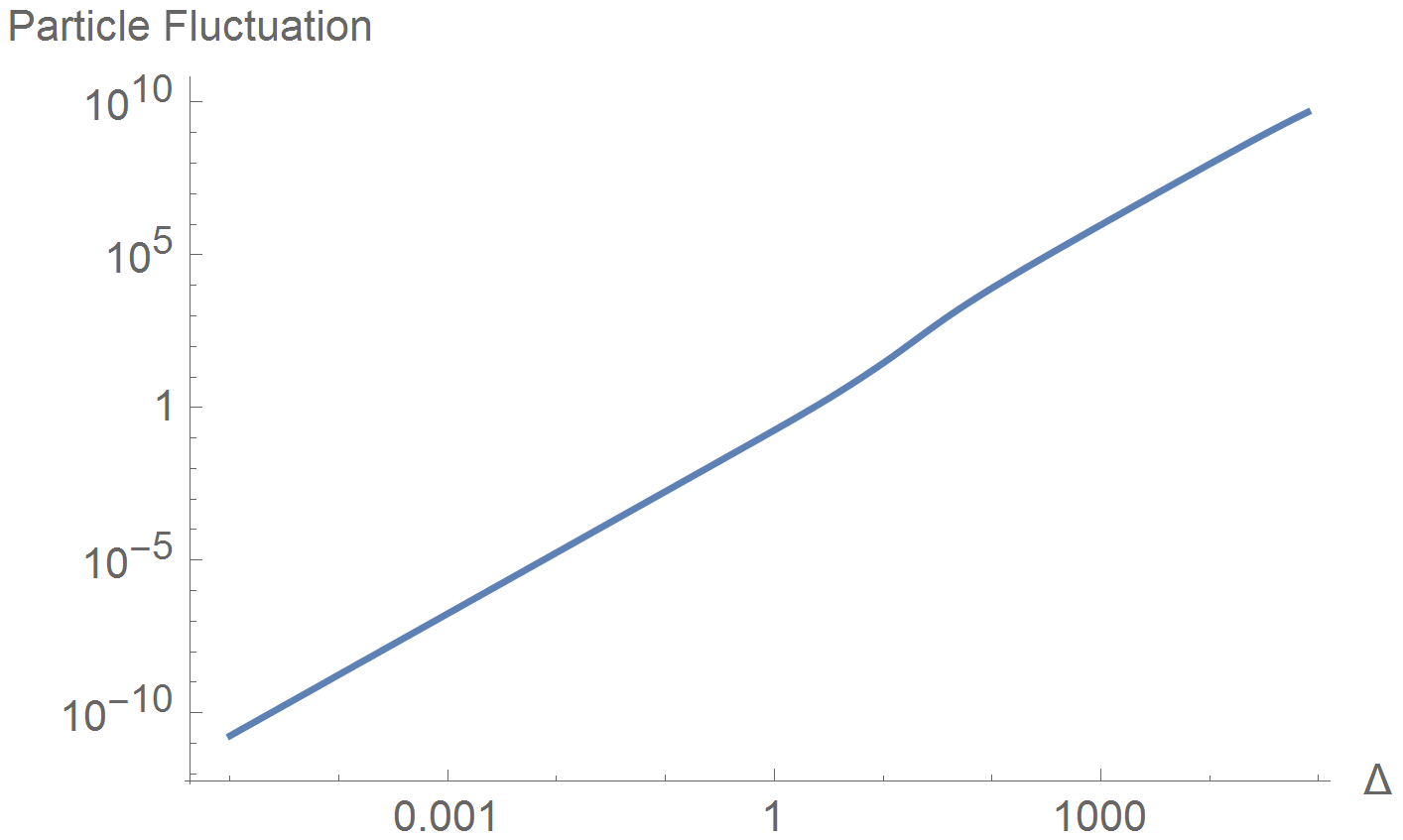}
\caption{The graph above is a plot of how the 0th order particle fluctuation, $\braket{\hat{N}_0}V_{1,0}'$, is affected by $\Delta$. We have utilized the following settings: $k_{wid}=10^6$}
\label{fig: Unitary}
\end{figure}


\begin{thebibliography}{99}

\bibitem{MOO70} G. T. Moore, J. Math. Phys. \textbf{11}, 2679(1970).
\bibitem{FUL76} S. A. Fulling and P. C. W. Davies, Proc. R. Soc. Lond.
A. \textbf{348}, 393(1976).
\bibitem{DAV77} P. C. W. Davies and S. A. Fulling, Proc. R. Soc. Lond.
A. \textbf{356}, 237(1977).
\bibitem{CAR87} R. D. Carlitz and R. S. Willey, Phys. Rev. D \textbf{36},
2327(1987).
\bibitem{HAW75} S. Hawking, Commun. Math. Phys. \textbf{43}, 199(1975).
\bibitem{OBA01} N. Obadia and R. Parentani, Phys. Rev. D \textbf{64},
044019(2001).
\bibitem{OBA03} N. Obadia and R. Parentani, Phys. Rev. D \textbf{67},
024022(2003).
\bibitem{MAR15} E. Martin-Martinez and J. Louko, Phys. Rev. Lett. \textbf{115},
031301(2015).
\bibitem{Daiqin2016}
		D. Su, C.T.M. Ho, R. B. Mann and T. C. Ralph, Quantum Circuit model for a uniformly accelerated mirror, arXiv:1602.02858
\bibitem{Daiqin2017}
		D. Su, T. C. Ralph, Decoherence of the radiation from an accelerated quantum source, arXiv:1705.07432
		\bibitem{PCWDavies1975}
		P.C.W. Davies, Scalar production in Schwarzschild and Rindler metrics, J. Phys. A \textbf{8}, 609(1975)		
		\bibitem{Unruh1976}
		W. G. Unruh, Phys. Rev. D \textbf{14},870(1976)

		\bibitem{Takagi1986}
		S. Takagi, Prog. Theor. Phys. Suppl. \textbf{88}, 1(1986)
		\bibitem{Crispino2008}
		L. Crispino, A. Higuchi and G. Matsas, Rev. Mod. Phys \textbf{80}, 787(2008)
		\bibitem{Fulling1973}
		S. A. Fulling, Phys. Rev. D \textbf{7}, 2850(1973)
		
		\bibitem{Birrell1982}
		N.D. Birrell and P.C.W. Davies, Quantum Fields in Curved Space(Cambridge University Press, Cambridge, England, 1982).
		\bibitem{Obadia2001}
		N. Obadia and R. Parentani, Phys. Rev. D \textbf{64}, 044019(2001)
		\bibitem{Lvovsky2009}
		A. I. Lvovsky and M. G. Raymer, Continuous-variable optical quantum-state tomography, Rev. Mod. Phys. \textbf{81}, 299(2009).
		\bibitem{Weedbrook2012}
		C. Weedbrook, S. Pirandola, R. Garcia-Patron, N. G. Cerf, T. C. Ralph, J.H Shapiro and S. Lloyd, Gaussian quantum information, Rev. Mod. Phys. \textbf{84}, 621 (2012)
		\bibitem{Scully1997}
		M. O. Scully and M.S. Zubairy, Quantum Optics (Cambridge University Press, Cambridge, 1997)
		\bibitem{Downes2013}
		T.G.Downes, T.C. Ralph and N. Walk, Phys. Rev. A \textbf{87}, 012327(2013)
		\bibitem{Rohde2007}
P. P. Rohde, W. Mauerer, C. Silberhorn, Spectural structure and decompositions of optical states, and their applications, New J. Phys. \textbf{9}, 91 (2007)
		\bibitem{Fulling1976}
S.A. Fulling and P. C. W. Davies, Proc. R. Soc. Lond. A. \textbf{348}, 393(1976)
		\bibitem{Grove1986}
P.G. Grove, Class. Quantum Grav. \textbf{3}, 193(1986)
		\bibitem{Frolov1979}
V.P. Frolov and E.M. Serebriany, J Phys. A: Math. Gen. \textbf{12}, 2415(1979)
		\bibitem{Frolov1980}
V.P. Frolov and E.M. Serebriany, J Phys. A: Math. Gen. \textbf{13}, 3205(1980)
		\bibitem{Frolov1999}
V.P. Frolov and D. Signh, Class. Quantum Grav. \textbf{16}, 3693(1999)
		\bibitem{Gravitation1973}
		C. W. Misner, K. Thorne, and J. A. Wheeler, Gravitation (Freeman, San Francisco, 1973)
		\bibitem{Davies1977}
		P.C.W. Davies and S. A. Fulling, Proc. R. Soc. Lond. A. \textbf{356}, 237(1977)
		\bibitem{Walker1985}
		W.R. Walker, PHYS. Rev. D \textbf{31}, 767(1985)
		\bibitem{Carlitz1987}
		R. D. Carlitz and R. S. Willey, Phys. Rev. D \textbf{36}, 2327(1987)
		\bibitem{Hawking1975}
		S. Hawking Commun. Math. Phys. \textbf{43}, 199 (1975)
		\bibitem{Hawking1976}
		S. Hawking, Phys. Rev. D \textbf{14}, 2460(1976)
		\bibitem{Susskind1993}
		L. Susskind, L. Thorlacius and J. Uglum, The stretched horizon and black hole complementarity, Phys. Rev. D \textbf{48}, 3743(1993)
		\bibitem{Stephens1994}
		C. R. Stephens, G. t'Hooft and B. G. Whitings, Black hole evaporation without information loss, Class. Quantum Grav. \textbf{11}, 621 (1994).
		\bibitem{Mathur2005}
		S. D. Mathur, The fuzzball proposal for black holes: an elementary review, Fortschr. Phys. \textbf{53}, 793 (2005).
		\bibitem{Hawking2016}
		S. W. Hawking, M. J. Perry and A. Strominger, Soft hair on black holes, Phys. Rev. Lett. \textbf{116}, 231301 (2016)
		\bibitem{VBaccetti2016}
		V. Baccetti, R.B. Mann and D. R. Terno, Role of evaporation in gravitational collapse, arXiv:1610.07839
		\bibitem{Braunstein2013}
		S. L. Braunstein and S. Pirandola, Better Late than Never: Information Retrieval from Black Holes, Phys. Rev. Lett. \textbf{11},101301(2013)
		\bibitem{Almheiri2013}
		A. Almheiri, D. Marolf, J.Polchinski and J.Sully, J. High Energy Phys. 02(2013) \textbf{062}
\bibitem{Nomura2016}
Nomura, Y. Salzetta, N. Why firewalls need not exist Physics Letters B, Elsevier BV, 2016, \textbf{761}, 62-69

\bibitem{Nomura2015a}
Nomura, Y., Sanches, F. Weinberg, S.J. J. High Energ. Phys. (2015) 2015: \textbf{158}. 

\bibitem{Nomura2015b}
Nomura, Y. Sanches, F. Weinberg, S. J. Black Hole Interior in Quantum Gravity Physical Review Letters, American Physical Society (APS), 2015, \textbf{114}

\bibitem{Subcycle}
C. Riek, P. Sulzer, M. Seeger, A. S. Moskalenko, G. Burkard, D. V. Seletskiy, A. Leitenstorfer, A. Subcycle quantum electrodynamics Nature, Springer Nature, 2017, \textbf{541}, 376-379

\bibitem{refractiveindex}
Yablonovitch. E, Accelerating reference frame for electromagnetic waves in a rapidly growing plasma: Unruh-Davies-Fulling-DeWitt radiation and the nonadiabatic Casimir effect Physical Review Letters, American Physical Society (APS), 1989, \textbf{62}, 1742-1745
\end{thebibliography}
\end{document}